\documentclass[longauth]{aa}

\usepackage{graphicx}
\usepackage{natbib}
\usepackage{scalerel}
\usepackage{amssymb}
\usepackage{float}
\usepackage{amsmath}
\usepackage{orcidlink}
\usepackage{supertabular}
\usepackage{pdflscape}

\usepackage{txfonts}
\hypersetup{
    colorlinks=true,
    linkcolor=blue,
    filecolor=magenta,      
    urlcolor=blue,
    citecolor=blue
}
\makeatletter
\renewcommand*\aa@pageof{, page \thepage{} of \pageref*{LastPage}}
\makeatother

\usepackage[utf8]{inputenc}

\usepackage[switch, modulo]{lineno}

\usepackage{euclid}

\begin{document}
%
%

\title{\Euclid preparation}

\subtitle{Far-infrared predictions for \Euclid galaxy catalogues: cluster, protocluster, and field}

\newcommand{\orcid}[1]{\orcidlink{#1}} 		   
\author{Euclid Collaboration: A.~Parmar\orcid{0009-0009-9390-9232}\thanks{\email{a.parmar21@imperial.ac.uk}}\inst{\ref{aff1}}
\and D.~L.~Clements\orcid{0000-0002-9548-5033}\inst{\ref{aff1}}
\and M.~Bolzonella\orcid{0000-0003-3278-4607}\inst{\ref{aff2}}
\and O.~Cucciati\orcid{0000-0002-9336-7551}\inst{\ref{aff2}}
\and L.~Pozzetti\orcid{0000-0001-7085-0412}\inst{\ref{aff2}}
\and H.~Dannerbauer\orcid{0000-0001-7147-3575}\inst{\ref{aff3}}
\and G.~Castignani\orcid{0000-0001-6831-0687}\inst{\ref{aff2}}
\and S.~Serjeant\orcid{0000-0002-0517-7943}\inst{\ref{aff4}}
\and L.~Wang\orcid{0000-0002-6736-9158}\inst{\ref{aff5},\ref{aff6}}
\and R.~Hill\inst{\ref{aff7}}
\and D.~Scott\orcid{0000-0002-6878-9840}\inst{\ref{aff7}}
\and J.~G.~Sorce\orcid{0000-0002-2307-2432}\inst{\ref{aff8},\ref{aff9}}
\and M.~Magliocchetti\orcid{0000-0001-9158-4838}\inst{\ref{aff10}}
\and F.~Pace\orcid{0000-0001-8039-0480}\inst{\ref{aff11},\ref{aff12},\ref{aff13}}
\and T.~T.~Thai\orcid{0000-0002-8408-4816}\inst{\ref{aff14}}
\and N.~Aghanim\orcid{0000-0002-6688-8992}\inst{\ref{aff9}}
\and B.~Altieri\orcid{0000-0003-3936-0284}\inst{\ref{aff15}}
\and S.~Andreon\orcid{0000-0002-2041-8784}\inst{\ref{aff16}}
\and N.~Auricchio\orcid{0000-0003-4444-8651}\inst{\ref{aff2}}
\and C.~Baccigalupi\orcid{0000-0002-8211-1630}\inst{\ref{aff17},\ref{aff18},\ref{aff19},\ref{aff20}}
\and M.~Baldi\orcid{0000-0003-4145-1943}\inst{\ref{aff21},\ref{aff2},\ref{aff22}}
\and S.~Bardelli\orcid{0000-0002-8900-0298}\inst{\ref{aff2}}
\and A.~Biviano\orcid{0000-0002-0857-0732}\inst{\ref{aff18},\ref{aff17}}
\and W.~Bon\inst{\ref{aff23}}
\and E.~Branchini\orcid{0000-0002-0808-6908}\inst{\ref{aff24},\ref{aff25},\ref{aff16}}
\and M.~Brescia\orcid{0000-0001-9506-5680}\inst{\ref{aff26},\ref{aff27}}
\and J.~Brinchmann\orcid{0000-0003-4359-8797}\inst{\ref{aff28},\ref{aff29},\ref{aff30}}
\and S.~Camera\orcid{0000-0003-3399-3574}\inst{\ref{aff11},\ref{aff12},\ref{aff13}}
\and G.~Ca\~nas-Herrera\orcid{0000-0003-2796-2149}\inst{\ref{aff31},\ref{aff32},\ref{aff33}}
\and V.~Capobianco\orcid{0000-0002-3309-7692}\inst{\ref{aff13}}
\and C.~Carbone\orcid{0000-0003-0125-3563}\inst{\ref{aff34}}
\and J.~Carretero\orcid{0000-0002-3130-0204}\inst{\ref{aff35},\ref{aff36}}
\and S.~Casas\orcid{0000-0002-4751-5138}\inst{\ref{aff37}}
\and M.~Castellano\orcid{0000-0001-9875-8263}\inst{\ref{aff38}}
\and S.~Cavuoti\orcid{0000-0002-3787-4196}\inst{\ref{aff27},\ref{aff39}}
\and A.~Cimatti\inst{\ref{aff40}}
\and C.~Colodro-Conde\inst{\ref{aff41}}
\and G.~Congedo\orcid{0000-0003-2508-0046}\inst{\ref{aff42}}
\and C.~J.~Conselice\orcid{0000-0003-1949-7638}\inst{\ref{aff43}}
\and L.~Conversi\orcid{0000-0002-6710-8476}\inst{\ref{aff44},\ref{aff15}}
\and Y.~Copin\orcid{0000-0002-5317-7518}\inst{\ref{aff45}}
\and F.~Courbin\orcid{0000-0003-0758-6510}\inst{\ref{aff46},\ref{aff47}}
\and H.~M.~Courtois\orcid{0000-0003-0509-1776}\inst{\ref{aff48}}
\and A.~Da~Silva\orcid{0000-0002-6385-1609}\inst{\ref{aff49},\ref{aff50}}
\and H.~Degaudenzi\orcid{0000-0002-5887-6799}\inst{\ref{aff51}}
\and G.~De~Lucia\orcid{0000-0002-6220-9104}\inst{\ref{aff18}}
\and H.~Dole\orcid{0000-0002-9767-3839}\inst{\ref{aff9}}
\and M.~Douspis\orcid{0000-0003-4203-3954}\inst{\ref{aff9}}
\and F.~Dubath\orcid{0000-0002-6533-2810}\inst{\ref{aff51}}
\and F.~Ducret\inst{\ref{aff23}}
\and C.~A.~J.~Duncan\orcid{0009-0003-3573-0791}\inst{\ref{aff42},\ref{aff43}}
\and X.~Dupac\inst{\ref{aff15}}
\and S.~Escoffier\orcid{0000-0002-2847-7498}\inst{\ref{aff52}}
\and M.~Farina\orcid{0000-0002-3089-7846}\inst{\ref{aff10}}
\and R.~Farinelli\inst{\ref{aff2}}
\and S.~Ferriol\inst{\ref{aff45}}
\and F.~Finelli\orcid{0000-0002-6694-3269}\inst{\ref{aff2},\ref{aff53}}
\and S.~Fotopoulou\orcid{0000-0002-9686-254X}\inst{\ref{aff54}}
\and M.~Frailis\orcid{0000-0002-7400-2135}\inst{\ref{aff18}}
\and E.~Franceschi\orcid{0000-0002-0585-6591}\inst{\ref{aff2}}
\and M.~Fumana\orcid{0000-0001-6787-5950}\inst{\ref{aff34}}
\and S.~Galeotta\orcid{0000-0002-3748-5115}\inst{\ref{aff18}}
\and K.~George\orcid{0000-0002-1734-8455}\inst{\ref{aff55}}
\and B.~Gillis\orcid{0000-0002-4478-1270}\inst{\ref{aff42}}
\and C.~Giocoli\orcid{0000-0002-9590-7961}\inst{\ref{aff2},\ref{aff22}}
\and J.~Gracia-Carpio\inst{\ref{aff56}}
\and A.~Grazian\orcid{0000-0002-5688-0663}\inst{\ref{aff57}}
\and F.~Grupp\inst{\ref{aff56},\ref{aff58}}
\and S.~V.~H.~Haugan\orcid{0000-0001-9648-7260}\inst{\ref{aff59}}
\and W.~Holmes\inst{\ref{aff60}}
\and F.~Hormuth\inst{\ref{aff61}}
\and A.~Hornstrup\orcid{0000-0002-3363-0936}\inst{\ref{aff62},\ref{aff63}}
\and K.~Jahnke\orcid{0000-0003-3804-2137}\inst{\ref{aff64}}
\and M.~Jhabvala\inst{\ref{aff65}}
\and E.~Keih\"anen\orcid{0000-0003-1804-7715}\inst{\ref{aff66}}
\and S.~Kermiche\orcid{0000-0002-0302-5735}\inst{\ref{aff52}}
\and A.~Kiessling\orcid{0000-0002-2590-1273}\inst{\ref{aff60}}
\and B.~Kubik\orcid{0009-0006-5823-4880}\inst{\ref{aff45}}
\and M.~K\"ummel\orcid{0000-0003-2791-2117}\inst{\ref{aff58}}
\and M.~Kunz\orcid{0000-0002-3052-7394}\inst{\ref{aff67}}
\and H.~Kurki-Suonio\orcid{0000-0002-4618-3063}\inst{\ref{aff68},\ref{aff69}}
\and A.~M.~C.~Le~Brun\orcid{0000-0002-0936-4594}\inst{\ref{aff70}}
\and S.~Ligori\orcid{0000-0003-4172-4606}\inst{\ref{aff13}}
\and P.~B.~Lilje\orcid{0000-0003-4324-7794}\inst{\ref{aff59}}
\and V.~Lindholm\orcid{0000-0003-2317-5471}\inst{\ref{aff68},\ref{aff69}}
\and I.~Lloro\orcid{0000-0001-5966-1434}\inst{\ref{aff71}}
\and G.~Mainetti\orcid{0000-0003-2384-2377}\inst{\ref{aff72}}
\and D.~Maino\inst{\ref{aff73},\ref{aff34},\ref{aff74}}
\and E.~Maiorano\orcid{0000-0003-2593-4355}\inst{\ref{aff2}}
\and O.~Mansutti\orcid{0000-0001-5758-4658}\inst{\ref{aff18}}
\and S.~Marcin\inst{\ref{aff75}}
\and O.~Marggraf\orcid{0000-0001-7242-3852}\inst{\ref{aff76}}
\and M.~Martinelli\orcid{0000-0002-6943-7732}\inst{\ref{aff38},\ref{aff77}}
\and N.~Martinet\orcid{0000-0003-2786-7790}\inst{\ref{aff23}}
\and F.~Marulli\orcid{0000-0002-8850-0303}\inst{\ref{aff78},\ref{aff2},\ref{aff22}}
\and R.~J.~Massey\orcid{0000-0002-6085-3780}\inst{\ref{aff79}}
\and S.~Maurogordato\inst{\ref{aff80}}
\and E.~Medinaceli\orcid{0000-0002-4040-7783}\inst{\ref{aff2}}
\and S.~Mei\orcid{0000-0002-2849-559X}\inst{\ref{aff81},\ref{aff82}}
\and M.~Melchior\inst{\ref{aff83}}
\and Y.~Mellier\thanks{Deceased}\inst{\ref{aff84},\ref{aff85}}
\and M.~Meneghetti\orcid{0000-0003-1225-7084}\inst{\ref{aff2},\ref{aff22}}
\and E.~Merlin\orcid{0000-0001-6870-8900}\inst{\ref{aff38}}
\and G.~Meylan\inst{\ref{aff86}}
\and A.~Mora\orcid{0000-0002-1922-8529}\inst{\ref{aff87}}
\and M.~Moresco\orcid{0000-0002-7616-7136}\inst{\ref{aff78},\ref{aff2}}
\and L.~Moscardini\orcid{0000-0002-3473-6716}\inst{\ref{aff78},\ref{aff2},\ref{aff22}}
\and R.~Nakajima\orcid{0009-0009-1213-7040}\inst{\ref{aff76}}
\and C.~Neissner\orcid{0000-0001-8524-4968}\inst{\ref{aff88},\ref{aff36}}
\and S.-M.~Niemi\orcid{0009-0005-0247-0086}\inst{\ref{aff31}}
\and C.~Padilla\orcid{0000-0001-7951-0166}\inst{\ref{aff88}}
\and S.~Paltani\orcid{0000-0002-8108-9179}\inst{\ref{aff51}}
\and F.~Pasian\orcid{0000-0002-4869-3227}\inst{\ref{aff18}}
\and K.~Pedersen\inst{\ref{aff89}}
\and V.~Pettorino\inst{\ref{aff31}}
\and S.~Pires\orcid{0000-0002-0249-2104}\inst{\ref{aff90}}
\and G.~Polenta\orcid{0000-0003-4067-9196}\inst{\ref{aff91}}
\and M.~Poncet\inst{\ref{aff92}}
\and L.~A.~Popa\inst{\ref{aff93}}
\and F.~Raison\orcid{0000-0002-7819-6918}\inst{\ref{aff56}}
\and R.~Rebolo\orcid{0000-0003-3767-7085}\inst{\ref{aff41},\ref{aff94},\ref{aff95}}
\and A.~Renzi\orcid{0000-0001-9856-1970}\inst{\ref{aff96},\ref{aff97}}
\and J.~Rhodes\orcid{0000-0002-4485-8549}\inst{\ref{aff60}}
\and G.~Riccio\inst{\ref{aff27}}
\and E.~Romelli\orcid{0000-0003-3069-9222}\inst{\ref{aff18}}
\and M.~Roncarelli\orcid{0000-0001-9587-7822}\inst{\ref{aff2}}
\and R.~Saglia\orcid{0000-0003-0378-7032}\inst{\ref{aff58},\ref{aff56}}
\and Z.~Sakr\orcid{0000-0002-4823-3757}\inst{\ref{aff98},\ref{aff99},\ref{aff100}}
\and A.~G.~S\'anchez\orcid{0000-0003-1198-831X}\inst{\ref{aff56}}
\and D.~Sapone\orcid{0000-0001-7089-4503}\inst{\ref{aff101}}
\and B.~Sartoris\orcid{0000-0003-1337-5269}\inst{\ref{aff58},\ref{aff18}}
\and P.~Schneider\orcid{0000-0001-8561-2679}\inst{\ref{aff76}}
\and T.~Schrabback\orcid{0000-0002-6987-7834}\inst{\ref{aff102}}
\and A.~Secroun\orcid{0000-0003-0505-3710}\inst{\ref{aff52}}
\and E.~Sefusatti\orcid{0000-0003-0473-1567}\inst{\ref{aff18},\ref{aff17},\ref{aff19}}
\and G.~Seidel\orcid{0000-0003-2907-353X}\inst{\ref{aff64}}
\and M.~Seiffert\orcid{0000-0002-7536-9393}\inst{\ref{aff60}}
\and S.~Serrano\orcid{0000-0002-0211-2861}\inst{\ref{aff103},\ref{aff104},\ref{aff105}}
\and P.~Simon\inst{\ref{aff76}}
\and C.~Sirignano\orcid{0000-0002-0995-7146}\inst{\ref{aff96},\ref{aff97}}
\and G.~Sirri\orcid{0000-0003-2626-2853}\inst{\ref{aff22}}
\and L.~Stanco\orcid{0000-0002-9706-5104}\inst{\ref{aff97}}
\and J.~Steinwagner\orcid{0000-0001-7443-1047}\inst{\ref{aff56}}
\and P.~Tallada-Cresp\'{i}\orcid{0000-0002-1336-8328}\inst{\ref{aff35},\ref{aff36}}
\and A.~N.~Taylor\inst{\ref{aff42}}
\and H.~I.~Teplitz\orcid{0000-0002-7064-5424}\inst{\ref{aff106}}
\and I.~Tereno\orcid{0000-0002-4537-6218}\inst{\ref{aff49},\ref{aff107}}
\and N.~Tessore\orcid{0000-0002-9696-7931}\inst{\ref{aff108},\ref{aff109}}
\and S.~Toft\orcid{0000-0003-3631-7176}\inst{\ref{aff110},\ref{aff111}}
\and R.~Toledo-Moreo\orcid{0000-0002-2997-4859}\inst{\ref{aff112}}
\and F.~Torradeflot\orcid{0000-0003-1160-1517}\inst{\ref{aff36},\ref{aff35}}
\and I.~Tutusaus\orcid{0000-0002-3199-0399}\inst{\ref{aff105},\ref{aff103},\ref{aff99}}
\and L.~Valenziano\orcid{0000-0002-1170-0104}\inst{\ref{aff2},\ref{aff53}}
\and J.~Valiviita\orcid{0000-0001-6225-3693}\inst{\ref{aff68},\ref{aff69}}
\and T.~Vassallo\orcid{0000-0001-6512-6358}\inst{\ref{aff18}}
\and A.~Veropalumbo\orcid{0000-0003-2387-1194}\inst{\ref{aff16},\ref{aff25},\ref{aff24}}
\and Y.~Wang\orcid{0000-0002-4749-2984}\inst{\ref{aff106}}
\and J.~Weller\orcid{0000-0002-8282-2010}\inst{\ref{aff58},\ref{aff56}}
\and G.~Zamorani\orcid{0000-0002-2318-301X}\inst{\ref{aff2}}
\and F.~M.~Zerbi\inst{\ref{aff16}}
\and E.~Zucca\orcid{0000-0002-5845-8132}\inst{\ref{aff2}}
\and V.~Allevato\orcid{0000-0001-7232-5152}\inst{\ref{aff27}}
\and M.~Ballardini\orcid{0000-0003-4481-3559}\inst{\ref{aff113},\ref{aff114},\ref{aff2}}
\and E.~Bozzo\orcid{0000-0002-8201-1525}\inst{\ref{aff51}}
\and C.~Burigana\orcid{0000-0002-3005-5796}\inst{\ref{aff115},\ref{aff53}}
\and R.~Cabanac\orcid{0000-0001-6679-2600}\inst{\ref{aff99}}
\and A.~Cappi\inst{\ref{aff2},\ref{aff80}}
\and D.~Di~Ferdinando\inst{\ref{aff22}}
\and J.~A.~Escartin~Vigo\inst{\ref{aff56}}
\and L.~Gabarra\orcid{0000-0002-8486-8856}\inst{\ref{aff116}}
\and W.~G.~Hartley\inst{\ref{aff51}}
\and S.~Matthew\orcid{0000-0001-8448-1697}\inst{\ref{aff42}}
\and M.~Maturi\orcid{0000-0002-3517-2422}\inst{\ref{aff98},\ref{aff117}}
\and N.~Mauri\orcid{0000-0001-8196-1548}\inst{\ref{aff40},\ref{aff22}}
\and R.~B.~Metcalf\orcid{0000-0003-3167-2574}\inst{\ref{aff78},\ref{aff2}}
\and A.~Pezzotta\orcid{0000-0003-0726-2268}\inst{\ref{aff16}}
\and M.~P\"ontinen\orcid{0000-0001-5442-2530}\inst{\ref{aff68}}
\and C.~Porciani\orcid{0000-0002-7797-2508}\inst{\ref{aff76}}
\and I.~Risso\orcid{0000-0003-2525-7761}\inst{\ref{aff16},\ref{aff25}}
\and V.~Scottez\orcid{0009-0008-3864-940X}\inst{\ref{aff84},\ref{aff118}}
\and M.~Sereno\orcid{0000-0003-0302-0325}\inst{\ref{aff2},\ref{aff22}}
\and M.~Tenti\orcid{0000-0002-4254-5901}\inst{\ref{aff22}}
\and M.~Viel\orcid{0000-0002-2642-5707}\inst{\ref{aff17},\ref{aff18},\ref{aff20},\ref{aff19},\ref{aff119}}
\and M.~Wiesmann\orcid{0009-0000-8199-5860}\inst{\ref{aff59}}
\and Y.~Akrami\orcid{0000-0002-2407-7956}\inst{\ref{aff120},\ref{aff121}}
\and S.~Anselmi\orcid{0000-0002-3579-9583}\inst{\ref{aff97},\ref{aff96},\ref{aff122}}
\and M.~Archidiacono\orcid{0000-0003-4952-9012}\inst{\ref{aff73},\ref{aff74}}
\and F.~Atrio-Barandela\orcid{0000-0002-2130-2513}\inst{\ref{aff123}}
\and P.~Bergamini\orcid{0000-0003-1383-9414}\inst{\ref{aff2}}
\and D.~Bertacca\orcid{0000-0002-2490-7139}\inst{\ref{aff96},\ref{aff57},\ref{aff97}}
\and M.~Bethermin\orcid{0000-0002-3915-2015}\inst{\ref{aff124}}
\and A.~Blanchard\orcid{0000-0001-8555-9003}\inst{\ref{aff99}}
\and L.~Blot\orcid{0000-0002-9622-7167}\inst{\ref{aff125},\ref{aff70}}
\and H.~B\"ohringer\orcid{0000-0001-8241-4204}\inst{\ref{aff56},\ref{aff55},\ref{aff126}}
\and M.~Bonici\orcid{0000-0002-8430-126X}\inst{\ref{aff127},\ref{aff34}}
\and S.~Borgani\orcid{0000-0001-6151-6439}\inst{\ref{aff128},\ref{aff17},\ref{aff18},\ref{aff19},\ref{aff119}}
\and M.~L.~Brown\orcid{0000-0002-0370-8077}\inst{\ref{aff43}}
\and S.~Bruton\orcid{0000-0002-6503-5218}\inst{\ref{aff129}}
\and A.~Calabro\orcid{0000-0003-2536-1614}\inst{\ref{aff38}}
\and B.~Camacho~Quevedo\orcid{0000-0002-8789-4232}\inst{\ref{aff17},\ref{aff20},\ref{aff18}}
\and F.~Caro\inst{\ref{aff38}}
\and C.~S.~Carvalho\inst{\ref{aff107}}
\and T.~Castro\orcid{0000-0002-6292-3228}\inst{\ref{aff18},\ref{aff19},\ref{aff17},\ref{aff119}}
\and F.~Cogato\orcid{0000-0003-4632-6113}\inst{\ref{aff78},\ref{aff2}}
\and S.~Conseil\orcid{0000-0002-3657-4191}\inst{\ref{aff45}}
\and A.~R.~Cooray\orcid{0000-0002-3892-0190}\inst{\ref{aff130}}
\and S.~Davini\orcid{0000-0003-3269-1718}\inst{\ref{aff25}}
\and G.~Desprez\orcid{0000-0001-8325-1742}\inst{\ref{aff6}}
\and A.~D\'iaz-S\'anchez\orcid{0000-0003-0748-4768}\inst{\ref{aff131}}
\and J.~J.~Diaz\orcid{0000-0003-2101-1078}\inst{\ref{aff41}}
\and S.~Di~Domizio\orcid{0000-0003-2863-5895}\inst{\ref{aff24},\ref{aff25}}
\and J.~M.~Diego\orcid{0000-0001-9065-3926}\inst{\ref{aff132}}
\and P.~Dimauro\orcid{0000-0001-7399-2854}\inst{\ref{aff133},\ref{aff38}}
\and M.~Y.~Elkhashab\orcid{0000-0001-9306-2603}\inst{\ref{aff18},\ref{aff19},\ref{aff128},\ref{aff17}}
\and A.~Enia\orcid{0000-0002-0200-2857}\inst{\ref{aff21},\ref{aff2}}
\and Y.~Fang\inst{\ref{aff58}}
\and A.~G.~Ferrari\orcid{0009-0005-5266-4110}\inst{\ref{aff22}}
\and A.~Finoguenov\orcid{0000-0002-4606-5403}\inst{\ref{aff68}}
\and A.~Fontana\orcid{0000-0003-3820-2823}\inst{\ref{aff38}}
\and F.~Fontanot\orcid{0000-0003-4744-0188}\inst{\ref{aff18},\ref{aff17}}
\and A.~Franco\orcid{0000-0002-4761-366X}\inst{\ref{aff134},\ref{aff135},\ref{aff136}}
\and K.~Ganga\orcid{0000-0001-8159-8208}\inst{\ref{aff81}}
\and J.~Garc\'ia-Bellido\orcid{0000-0002-9370-8360}\inst{\ref{aff120}}
\and T.~Gasparetto\orcid{0000-0002-7913-4866}\inst{\ref{aff38}}
\and V.~Gautard\inst{\ref{aff137}}
\and E.~Gaztanaga\orcid{0000-0001-9632-0815}\inst{\ref{aff105},\ref{aff103},\ref{aff138}}
\and F.~Giacomini\orcid{0000-0002-3129-2814}\inst{\ref{aff22}}
\and F.~Gianotti\orcid{0000-0003-4666-119X}\inst{\ref{aff2}}
\and G.~Gozaliasl\orcid{0000-0002-0236-919X}\inst{\ref{aff139},\ref{aff68}}
\and M.~Guidi\orcid{0000-0001-9408-1101}\inst{\ref{aff21},\ref{aff2}}
\and C.~M.~Gutierrez\orcid{0000-0001-7854-783X}\inst{\ref{aff3}}
\and A.~Hall\orcid{0000-0002-3139-8651}\inst{\ref{aff42}}
\and S.~Hemmati\orcid{0000-0003-2226-5395}\inst{\ref{aff140}}
\and H.~Hildebrandt\orcid{0000-0002-9814-3338}\inst{\ref{aff141}}
\and J.~Hjorth\orcid{0000-0002-4571-2306}\inst{\ref{aff89}}
\and J.~J.~E.~Kajava\orcid{0000-0002-3010-8333}\inst{\ref{aff142},\ref{aff143}}
\and Y.~Kang\orcid{0009-0000-8588-7250}\inst{\ref{aff51}}
\and V.~Kansal\orcid{0000-0002-4008-6078}\inst{\ref{aff144},\ref{aff145}}
\and D.~Karagiannis\orcid{0000-0002-4927-0816}\inst{\ref{aff113},\ref{aff146}}
\and K.~Kiiveri\inst{\ref{aff66}}
\and J.~Kim\orcid{0000-0003-2776-2761}\inst{\ref{aff116}}
\and C.~C.~Kirkpatrick\inst{\ref{aff66}}
\and S.~Kruk\orcid{0000-0001-8010-8879}\inst{\ref{aff15}}
\and J.~Le~Graet\orcid{0000-0001-6523-7971}\inst{\ref{aff52}}
\and L.~Legrand\orcid{0000-0003-0610-5252}\inst{\ref{aff147},\ref{aff148}}
\and M.~Lembo\orcid{0000-0002-5271-5070}\inst{\ref{aff85},\ref{aff113},\ref{aff114}}
\and F.~Lepori\orcid{0009-0000-5061-7138}\inst{\ref{aff149}}
\and G.~Leroy\orcid{0009-0004-2523-4425}\inst{\ref{aff150},\ref{aff79}}
\and G.~F.~Lesci\orcid{0000-0002-4607-2830}\inst{\ref{aff78},\ref{aff2}}
\and J.~Lesgourgues\orcid{0000-0001-7627-353X}\inst{\ref{aff37}}
\and T.~I.~Liaudat\orcid{0000-0002-9104-314X}\inst{\ref{aff151}}
\and A.~Loureiro\orcid{0000-0002-4371-0876}\inst{\ref{aff152},\ref{aff1}}
\and J.~Macias-Perez\orcid{0000-0002-5385-2763}\inst{\ref{aff153}}
\and G.~Maggio\orcid{0000-0003-4020-4836}\inst{\ref{aff18}}
\and C.~Mancini\orcid{0000-0002-4297-0561}\inst{\ref{aff34}}
\and F.~Mannucci\orcid{0000-0002-4803-2381}\inst{\ref{aff154}}
\and R.~Maoli\orcid{0000-0002-6065-3025}\inst{\ref{aff155},\ref{aff38}}
\and C.~J.~A.~P.~Martins\orcid{0000-0002-4886-9261}\inst{\ref{aff156},\ref{aff28}}
\and L.~Maurin\orcid{0000-0002-8406-0857}\inst{\ref{aff9}}
\and M.~Miluzio\inst{\ref{aff15},\ref{aff157}}
\and P.~Monaco\orcid{0000-0003-2083-7564}\inst{\ref{aff128},\ref{aff18},\ref{aff19},\ref{aff17},\ref{aff119}}
\and C.~Moretti\orcid{0000-0003-3314-8936}\inst{\ref{aff18},\ref{aff17},\ref{aff19},\ref{aff20}}
\and G.~Morgante\inst{\ref{aff2}}
\and K.~Naidoo\orcid{0000-0002-9182-1802}\inst{\ref{aff138},\ref{aff109}}
\and P.~Natoli\orcid{0000-0003-0126-9100}\inst{\ref{aff113},\ref{aff114}}
\and A.~Navarro-Alsina\orcid{0000-0002-3173-2592}\inst{\ref{aff76}}
\and S.~Nesseris\orcid{0000-0002-0567-0324}\inst{\ref{aff120}}
\and D.~Paoletti\orcid{0000-0003-4761-6147}\inst{\ref{aff2},\ref{aff53}}
\and F.~Passalacqua\orcid{0000-0002-8606-4093}\inst{\ref{aff96},\ref{aff97}}
\and K.~Paterson\orcid{0000-0001-8340-3486}\inst{\ref{aff64}}
\and L.~Patrizii\inst{\ref{aff22}}
\and A.~Pisani\orcid{0000-0002-6146-4437}\inst{\ref{aff52}}
\and D.~Potter\orcid{0000-0002-0757-5195}\inst{\ref{aff149}}
\and S.~Quai\orcid{0000-0002-0449-8163}\inst{\ref{aff78},\ref{aff2}}
\and M.~Radovich\orcid{0000-0002-3585-866X}\inst{\ref{aff57}}
\and P.-F.~Rocci\inst{\ref{aff9}}
\and G.~Rodighiero\orcid{0000-0002-9415-2296}\inst{\ref{aff96},\ref{aff57}}
\and S.~Sacquegna\orcid{0000-0002-8433-6630}\inst{\ref{aff158},\ref{aff135},\ref{aff134}}
\and M.~Sahl\'en\orcid{0000-0003-0973-4804}\inst{\ref{aff159}}
\and D.~B.~Sanders\orcid{0000-0002-1233-9998}\inst{\ref{aff160}}
\and E.~Sarpa\orcid{0000-0002-1256-655X}\inst{\ref{aff20},\ref{aff119},\ref{aff19}}
\and A.~Schneider\orcid{0000-0001-7055-8104}\inst{\ref{aff149}}
\and D.~Sciotti\orcid{0009-0008-4519-2620}\inst{\ref{aff38},\ref{aff77}}
\and E.~Sellentin\inst{\ref{aff161},\ref{aff33}}
\and F.~Shankar\orcid{0000-0001-8973-5051}\inst{\ref{aff162}}
\and L.~C.~Smith\orcid{0000-0002-3259-2771}\inst{\ref{aff163}}
\and K.~Tanidis\orcid{0000-0001-9843-5130}\inst{\ref{aff116}}
\and C.~Tao\orcid{0000-0001-7961-8177}\inst{\ref{aff52}}
\and G.~Testera\inst{\ref{aff25}}
\and R.~Teyssier\orcid{0000-0001-7689-0933}\inst{\ref{aff164}}
\and S.~Tosi\orcid{0000-0002-7275-9193}\inst{\ref{aff24},\ref{aff25},\ref{aff16}}
\and A.~Troja\orcid{0000-0003-0239-4595}\inst{\ref{aff96},\ref{aff97}}
\and M.~Tucci\inst{\ref{aff51}}
\and C.~Valieri\inst{\ref{aff22}}
\and A.~Venhola\orcid{0000-0001-6071-4564}\inst{\ref{aff165}}
\and D.~Vergani\orcid{0000-0003-0898-2216}\inst{\ref{aff2}}
\and G.~Verza\orcid{0000-0002-1886-8348}\inst{\ref{aff166}}
\and P.~Vielzeuf\orcid{0000-0003-2035-9339}\inst{\ref{aff52}}
\and N.~A.~Walton\orcid{0000-0003-3983-8778}\inst{\ref{aff163}}}
										   
\institute{Astrophysics Group, Blackett Laboratory, Imperial College London, London SW7 2AZ, UK\label{aff1}
\and
INAF-Osservatorio di Astrofisica e Scienza dello Spazio di Bologna, Via Piero Gobetti 93/3, 40129 Bologna, Italy\label{aff2}
\and
 Instituto de Astrof\'{\i}sica de Canarias, E-38205 La Laguna; Universidad de La Laguna, Dpto. Astrof\'\i sica, E-38206 La Laguna, Tenerife, Spain\label{aff3}
\and
School of Physical Sciences, The Open University, Milton Keynes, MK7 6AA, UK\label{aff4}
\and
SRON Netherlands Institute for Space Research, Landleven 12, 9747 AD, Groningen, The Netherlands\label{aff5}
\and
Kapteyn Astronomical Institute, University of Groningen, PO Box 800, 9700 AV Groningen, The Netherlands\label{aff6}
\and
Department of Physics and Astronomy, University of British Columbia, Vancouver, BC V6T 1Z1, Canada\label{aff7}
\and
Univ. Lille, CNRS, Centrale Lille, UMR 9189 CRIStAL, 59000 Lille, France\label{aff8}
\and
Universit\'e Paris-Saclay, CNRS, Institut d'astrophysique spatiale, 91405, Orsay, France\label{aff9}
\and
INAF-Istituto di Astrofisica e Planetologia Spaziali, via del Fosso del Cavaliere, 100, 00100 Roma, Italy\label{aff10}
\and
Dipartimento di Fisica, Universit\`a degli Studi di Torino, Via P. Giuria 1, 10125 Torino, Italy\label{aff11}
\and
INFN-Sezione di Torino, Via P. Giuria 1, 10125 Torino, Italy\label{aff12}
\and
INAF-Osservatorio Astrofisico di Torino, Via Osservatorio 20, 10025 Pino Torinese (TO), Italy\label{aff13}
\and
National Astronomical Observatory of Japan, 2-21-1 Osawa, Mitaka, Tokyo 181-8588, Japan\label{aff14}
\and
ESAC/ESA, Camino Bajo del Castillo, s/n., Urb. Villafranca del Castillo, 28692 Villanueva de la Ca\~nada, Madrid, Spain\label{aff15}
\and
INAF-Osservatorio Astronomico di Brera, Via Brera 28, 20122 Milano, Italy\label{aff16}
\and
IFPU, Institute for Fundamental Physics of the Universe, via Beirut 2, 34151 Trieste, Italy\label{aff17}
\and
INAF-Osservatorio Astronomico di Trieste, Via G. B. Tiepolo 11, 34143 Trieste, Italy\label{aff18}
\and
INFN, Sezione di Trieste, Via Valerio 2, 34127 Trieste TS, Italy\label{aff19}
\and
SISSA, International School for Advanced Studies, Via Bonomea 265, 34136 Trieste TS, Italy\label{aff20}
\and
Dipartimento di Fisica e Astronomia, Universit\`a di Bologna, Via Gobetti 93/2, 40129 Bologna, Italy\label{aff21}
\and
INFN-Sezione di Bologna, Viale Berti Pichat 6/2, 40127 Bologna, Italy\label{aff22}
\and
Aix-Marseille Universit\'e, CNRS, CNES, LAM, Marseille, France\label{aff23}
\and
Dipartimento di Fisica, Universit\`a di Genova, Via Dodecaneso 33, 16146, Genova, Italy\label{aff24}
\and
INFN-Sezione di Genova, Via Dodecaneso 33, 16146, Genova, Italy\label{aff25}
\and
Department of Physics "E. Pancini", University Federico II, Via Cinthia 6, 80126, Napoli, Italy\label{aff26}
\and
INAF-Osservatorio Astronomico di Capodimonte, Via Moiariello 16, 80131 Napoli, Italy\label{aff27}
\and
Instituto de Astrof\'isica e Ci\^encias do Espa\c{c}o, Universidade do Porto, CAUP, Rua das Estrelas, PT4150-762 Porto, Portugal\label{aff28}
\and
Faculdade de Ci\^encias da Universidade do Porto, Rua do Campo de Alegre, 4150-007 Porto, Portugal\label{aff29}
\and
European Southern Observatory, Karl-Schwarzschild-Str.~2, 85748 Garching, Germany\label{aff30}
\and
European Space Agency/ESTEC, Keplerlaan 1, 2201 AZ Noordwijk, The Netherlands\label{aff31}
\and
Institute Lorentz, Leiden University, Niels Bohrweg 2, 2333 CA Leiden, The Netherlands\label{aff32}
\and
Leiden Observatory, Leiden University, Einsteinweg 55, 2333 CC Leiden, The Netherlands\label{aff33}
\and
INAF-IASF Milano, Via Alfonso Corti 12, 20133 Milano, Italy\label{aff34}
\and
Centro de Investigaciones Energ\'eticas, Medioambientales y Tecnol\'ogicas (CIEMAT), Avenida Complutense 40, 28040 Madrid, Spain\label{aff35}
\and
Port d'Informaci\'{o} Cient\'{i}fica, Campus UAB, C. Albareda s/n, 08193 Bellaterra (Barcelona), Spain\label{aff36}
\and
Institute for Theoretical Particle Physics and Cosmology (TTK), RWTH Aachen University, 52056 Aachen, Germany\label{aff37}
\and
INAF-Osservatorio Astronomico di Roma, Via Frascati 33, 00078 Monteporzio Catone, Italy\label{aff38}
\and
INFN section of Naples, Via Cinthia 6, 80126, Napoli, Italy\label{aff39}
\and
Dipartimento di Fisica e Astronomia "Augusto Righi" - Alma Mater Studiorum Universit\`a di Bologna, Viale Berti Pichat 6/2, 40127 Bologna, Italy\label{aff40}
\and
Instituto de Astrof\'{\i}sica de Canarias, E-38205 La Laguna, Tenerife, Spain\label{aff41}
\and
Institute for Astronomy, University of Edinburgh, Royal Observatory, Blackford Hill, Edinburgh EH9 3HJ, UK\label{aff42}
\and
Jodrell Bank Centre for Astrophysics, Department of Physics and Astronomy, University of Manchester, Oxford Road, Manchester M13 9PL, UK\label{aff43}
\and
European Space Agency/ESRIN, Largo Galileo Galilei 1, 00044 Frascati, Roma, Italy\label{aff44}
\and
Universit\'e Claude Bernard Lyon 1, CNRS/IN2P3, IP2I Lyon, UMR 5822, Villeurbanne, F-69100, France\label{aff45}
\and
Institut de Ci\`{e}ncies del Cosmos (ICCUB), Universitat de Barcelona (IEEC-UB), Mart\'{i} i Franqu\`{e}s 1, 08028 Barcelona, Spain\label{aff46}
\and
Instituci\'o Catalana de Recerca i Estudis Avan\c{c}ats (ICREA), Passeig de Llu\'{\i}s Companys 23, 08010 Barcelona, Spain\label{aff47}
\and
UCB Lyon 1, CNRS/IN2P3, IUF, IP2I Lyon, 4 rue Enrico Fermi, 69622 Villeurbanne, France\label{aff48}
\and
Departamento de F\'isica, Faculdade de Ci\^encias, Universidade de Lisboa, Edif\'icio C8, Campo Grande, PT1749-016 Lisboa, Portugal\label{aff49}
\and
Instituto de Astrof\'isica e Ci\^encias do Espa\c{c}o, Faculdade de Ci\^encias, Universidade de Lisboa, Campo Grande, 1749-016 Lisboa, Portugal\label{aff50}
\and
Department of Astronomy, University of Geneva, ch. d'Ecogia 16, 1290 Versoix, Switzerland\label{aff51}
\and
Aix-Marseille Universit\'e, CNRS/IN2P3, CPPM, Marseille, France\label{aff52}
\and
INFN-Bologna, Via Irnerio 46, 40126 Bologna, Italy\label{aff53}
\and
School of Physics, HH Wills Physics Laboratory, University of Bristol, Tyndall Avenue, Bristol, BS8 1TL, UK\label{aff54}
\and
University Observatory, LMU Faculty of Physics, Scheinerstr.~1, 81679 Munich, Germany\label{aff55}
\and
Max Planck Institute for Extraterrestrial Physics, Giessenbachstr. 1, 85748 Garching, Germany\label{aff56}
\and
INAF-Osservatorio Astronomico di Padova, Via dell'Osservatorio 5, 35122 Padova, Italy\label{aff57}
\and
Universit\"ats-Sternwarte M\"unchen, Fakult\"at f\"ur Physik, Ludwig-Maximilians-Universit\"at M\"unchen, Scheinerstr.~1, 81679 M\"unchen, Germany\label{aff58}
\and
Institute of Theoretical Astrophysics, University of Oslo, P.O. Box 1029 Blindern, 0315 Oslo, Norway\label{aff59}
\and
Jet Propulsion Laboratory, California Institute of Technology, 4800 Oak Grove Drive, Pasadena, CA, 91109, USA\label{aff60}
\and
Felix Hormuth Engineering, Goethestr. 17, 69181 Leimen, Germany\label{aff61}
\and
Technical University of Denmark, Elektrovej 327, 2800 Kgs. Lyngby, Denmark\label{aff62}
\and
Cosmic Dawn Center (DAWN), Denmark\label{aff63}
\and
Max-Planck-Institut f\"ur Astronomie, K\"onigstuhl 17, 69117 Heidelberg, Germany\label{aff64}
\and
NASA Goddard Space Flight Center, Greenbelt, MD 20771, USA\label{aff65}
\and
Department of Physics and Helsinki Institute of Physics, Gustaf H\"allstr\"omin katu 2, University of Helsinki, 00014 Helsinki, Finland\label{aff66}
\and
Universit\'e de Gen\`eve, D\'epartement de Physique Th\'eorique and Centre for Astroparticle Physics, 24 quai Ernest-Ansermet, CH-1211 Gen\`eve 4, Switzerland\label{aff67}
\and
Department of Physics, P.O. Box 64, University of Helsinki, 00014 Helsinki, Finland\label{aff68}
\and
Helsinki Institute of Physics, Gustaf H{\"a}llstr{\"o}min katu 2, University of Helsinki, 00014 Helsinki, Finland\label{aff69}
\and
Laboratoire d'etude de l'Univers et des phenomenes eXtremes, Observatoire de Paris, Universit\'e PSL, Sorbonne Universit\'e, CNRS, 92190 Meudon, France\label{aff70}
\and
SKAO, Jodrell Bank, Lower Withington, Macclesfield SK11 9FT, UK\label{aff71}
\and
Centre de Calcul de l'IN2P3/CNRS, 21 avenue Pierre de Coubertin 69627 Villeurbanne Cedex, France\label{aff72}
\and
Dipartimento di Fisica "Aldo Pontremoli", Universit\`a degli Studi di Milano, Via Celoria 16, 20133 Milano, Italy\label{aff73}
\and
INFN-Sezione di Milano, Via Celoria 16, 20133 Milano, Italy\label{aff74}
\and
University of Applied Sciences and Arts of Northwestern Switzerland, School of Computer Science, 5210 Windisch, Switzerland\label{aff75}
\and
Universit\"at Bonn, Argelander-Institut f\"ur Astronomie, Auf dem H\"ugel 71, 53121 Bonn, Germany\label{aff76}
\and
INFN-Sezione di Roma, Piazzale Aldo Moro, 2 - c/o Dipartimento di Fisica, Edificio G. Marconi, 00185 Roma, Italy\label{aff77}
\and
Dipartimento di Fisica e Astronomia "Augusto Righi" - Alma Mater Studiorum Universit\`a di Bologna, via Piero Gobetti 93/2, 40129 Bologna, Italy\label{aff78}
\and
Department of Physics, Institute for Computational Cosmology, Durham University, South Road, Durham, DH1 3LE, UK\label{aff79}
\and
Universit\'e C\^{o}te d'Azur, Observatoire de la C\^{o}te d'Azur, CNRS, Laboratoire Lagrange, Bd de l'Observatoire, CS 34229, 06304 Nice cedex 4, France\label{aff80}
\and
Universit\'e Paris Cit\'e, CNRS, Astroparticule et Cosmologie, 75013 Paris, France\label{aff81}
\and
CNRS-UCB International Research Laboratory, Centre Pierre Bin\'etruy, IRL2007, CPB-IN2P3, Berkeley, USA\label{aff82}
\and
University of Applied Sciences and Arts of Northwestern Switzerland, School of Engineering, 5210 Windisch, Switzerland\label{aff83}
\and
Institut d'Astrophysique de Paris, 98bis Boulevard Arago, 75014, Paris, France\label{aff84}
\and
Institut d'Astrophysique de Paris, UMR 7095, CNRS, and Sorbonne Universit\'e, 98 bis boulevard Arago, 75014 Paris, France\label{aff85}
\and
Institute of Physics, Laboratory of Astrophysics, Ecole Polytechnique F\'ed\'erale de Lausanne (EPFL), Observatoire de Sauverny, 1290 Versoix, Switzerland\label{aff86}
\and
Telespazio UK S.L. for European Space Agency (ESA), Camino bajo del Castillo, s/n, Urbanizacion Villafranca del Castillo, Villanueva de la Ca\~nada, 28692 Madrid, Spain\label{aff87}
\and
Institut de F\'{i}sica d'Altes Energies (IFAE), The Barcelona Institute of Science and Technology, Campus UAB, 08193 Bellaterra (Barcelona), Spain\label{aff88}
\and
DARK, Niels Bohr Institute, University of Copenhagen, Jagtvej 155, 2200 Copenhagen, Denmark\label{aff89}
\and
Universit\'e Paris-Saclay, Universit\'e Paris Cit\'e, CEA, CNRS, AIM, 91191, Gif-sur-Yvette, France\label{aff90}
\and
Space Science Data Center, Italian Space Agency, via del Politecnico snc, 00133 Roma, Italy\label{aff91}
\and
Centre National d'Etudes Spatiales -- Centre spatial de Toulouse, 18 avenue Edouard Belin, 31401 Toulouse Cedex 9, France\label{aff92}
\and
Institute of Space Science, Str. Atomistilor, nr. 409 M\u{a}gurele, Ilfov, 077125, Romania\label{aff93}
\and
Consejo Superior de Investigaciones Cientificas, Calle Serrano 117, 28006 Madrid, Spain\label{aff94}
\and
Universidad de La Laguna, Dpto. Astrof\'\i sica, E-38206 La Laguna, Tenerife, Spain\label{aff95}
\and
Dipartimento di Fisica e Astronomia "G. Galilei", Universit\`a di Padova, Via Marzolo 8, 35131 Padova, Italy\label{aff96}
\and
INFN-Padova, Via Marzolo 8, 35131 Padova, Italy\label{aff97}
\and
Institut f\"ur Theoretische Physik, University of Heidelberg, Philosophenweg 16, 69120 Heidelberg, Germany\label{aff98}
\and
Institut de Recherche en Astrophysique et Plan\'etologie (IRAP), Universit\'e de Toulouse, CNRS, UPS, CNES, 14 Av. Edouard Belin, 31400 Toulouse, France\label{aff99}
\and
Universit\'e St Joseph; Faculty of Sciences, Beirut, Lebanon\label{aff100}
\and
Departamento de F\'isica, FCFM, Universidad de Chile, Blanco Encalada 2008, Santiago, Chile\label{aff101}
\and
Universit\"at Innsbruck, Institut f\"ur Astro- und Teilchenphysik, Technikerstr. 25/8, 6020 Innsbruck, Austria\label{aff102}
\and
Institut d'Estudis Espacials de Catalunya (IEEC),  Edifici RDIT, Campus UPC, 08860 Castelldefels, Barcelona, Spain\label{aff103}
\and
Satlantis, University Science Park, Sede Bld 48940, Leioa-Bilbao, Spain\label{aff104}
\and
Institute of Space Sciences (ICE, CSIC), Campus UAB, Carrer de Can Magrans, s/n, 08193 Barcelona, Spain\label{aff105}
\and
Infrared Processing and Analysis Center, California Institute of Technology, Pasadena, CA 91125, USA\label{aff106}
\and
Instituto de Astrof\'isica e Ci\^encias do Espa\c{c}o, Faculdade de Ci\^encias, Universidade de Lisboa, Tapada da Ajuda, 1349-018 Lisboa, Portugal\label{aff107}
\and
Mullard Space Science Laboratory, University College London, Holmbury St Mary, Dorking, Surrey RH5 6NT, UK\label{aff108}
\and
Department of Physics and Astronomy, University College London, Gower Street, London WC1E 6BT, UK\label{aff109}
\and
Cosmic Dawn Center (DAWN)\label{aff110}
\and
Niels Bohr Institute, University of Copenhagen, Jagtvej 128, 2200 Copenhagen, Denmark\label{aff111}
\and
Universidad Polit\'ecnica de Cartagena, Departamento de Electr\'onica y Tecnolog\'ia de Computadoras,  Plaza del Hospital 1, 30202 Cartagena, Spain\label{aff112}
\and
Dipartimento di Fisica e Scienze della Terra, Universit\`a degli Studi di Ferrara, Via Giuseppe Saragat 1, 44122 Ferrara, Italy\label{aff113}
\and
Istituto Nazionale di Fisica Nucleare, Sezione di Ferrara, Via Giuseppe Saragat 1, 44122 Ferrara, Italy\label{aff114}
\and
INAF, Istituto di Radioastronomia, Via Piero Gobetti 101, 40129 Bologna, Italy\label{aff115}
\and
Department of Physics, Oxford University, Keble Road, Oxford OX1 3RH, UK\label{aff116}
\and
Zentrum f\"ur Astronomie, Universit\"at Heidelberg, Philosophenweg 12, 69120 Heidelberg, Germany\label{aff117}
\and
ICL, Junia, Universit\'e Catholique de Lille, LITL, 59000 Lille, France\label{aff118}
\and
ICSC - Centro Nazionale di Ricerca in High Performance Computing, Big Data e Quantum Computing, Via Magnanelli 2, Bologna, Italy\label{aff119}
\and
Instituto de F\'isica Te\'orica UAM-CSIC, Campus de Cantoblanco, 28049 Madrid, Spain\label{aff120}
\and
CERCA/ISO, Department of Physics, Case Western Reserve University, 10900 Euclid Avenue, Cleveland, OH 44106, USA\label{aff121}
\and
Laboratoire Univers et Th\'eorie, Observatoire de Paris, Universit\'e PSL, Universit\'e Paris Cit\'e, CNRS, 92190 Meudon, France\label{aff122}
\and
Departamento de F{\'\i}sica Fundamental. Universidad de Salamanca. Plaza de la Merced s/n. 37008 Salamanca, Spain\label{aff123}
\and
Universit\'e de Strasbourg, CNRS, Observatoire astronomique de Strasbourg, UMR 7550, 67000 Strasbourg, France\label{aff124}
\and
Center for Data-Driven Discovery, Kavli IPMU (WPI), UTIAS, The University of Tokyo, Kashiwa, Chiba 277-8583, Japan\label{aff125}
\and
Max-Planck-Institut f\"ur Physik, Boltzmannstr. 8, 85748 Garching, Germany\label{aff126}
\and
Waterloo Centre for Astrophysics, University of Waterloo, Waterloo, Ontario N2L 3G1, Canada\label{aff127}
\and
Dipartimento di Fisica - Sezione di Astronomia, Universit\`a di Trieste, Via Tiepolo 11, 34131 Trieste, Italy\label{aff128}
\and
California Institute of Technology, 1200 E California Blvd, Pasadena, CA 91125, USA\label{aff129}
\and
Department of Physics \& Astronomy, University of California Irvine, Irvine CA 92697, USA\label{aff130}
\and
Departamento F\'isica Aplicada, Universidad Polit\'ecnica de Cartagena, Campus Muralla del Mar, 30202 Cartagena, Murcia, Spain\label{aff131}
\and
Instituto de F\'isica de Cantabria, Edificio Juan Jord\'a, Avenida de los Castros, 39005 Santander, Spain\label{aff132}
\and
Observatorio Nacional, Rua General Jose Cristino, 77-Bairro Imperial de Sao Cristovao, Rio de Janeiro, 20921-400, Brazil\label{aff133}
\and
INFN, Sezione di Lecce, Via per Arnesano, CP-193, 73100, Lecce, Italy\label{aff134}
\and
Department of Mathematics and Physics E. De Giorgi, University of Salento, Via per Arnesano, CP-I93, 73100, Lecce, Italy\label{aff135}
\and
INAF-Sezione di Lecce, c/o Dipartimento Matematica e Fisica, Via per Arnesano, 73100, Lecce, Italy\label{aff136}
\and
CEA Saclay, DFR/IRFU, Service d'Astrophysique, Bat. 709, 91191 Gif-sur-Yvette, France\label{aff137}
\and
Institute of Cosmology and Gravitation, University of Portsmouth, Portsmouth PO1 3FX, UK\label{aff138}
\and
Department of Computer Science, Aalto University, PO Box 15400, Espoo, FI-00 076, Finland\label{aff139}
\and
Caltech/IPAC, 1200 E. California Blvd., Pasadena, CA 91125, USA\label{aff140}
\and
Ruhr University Bochum, Faculty of Physics and Astronomy, Astronomical Institute (AIRUB), German Centre for Cosmological Lensing (GCCL), 44780 Bochum, Germany\label{aff141}
\and
Department of Physics and Astronomy, Vesilinnantie 5, University of Turku, 20014 Turku, Finland\label{aff142}
\and
Serco for European Space Agency (ESA), Camino bajo del Castillo, s/n, Urbanizacion Villafranca del Castillo, Villanueva de la Ca\~nada, 28692 Madrid, Spain\label{aff143}
\and
ARC Centre of Excellence for Dark Matter Particle Physics, Melbourne, Australia\label{aff144}
\and
Centre for Astrophysics \& Supercomputing, Swinburne University of Technology,  Hawthorn, Victoria 3122, Australia\label{aff145}
\and
Department of Physics and Astronomy, University of the Western Cape, Bellville, Cape Town, 7535, South Africa\label{aff146}
\and
DAMTP, Centre for Mathematical Sciences, Wilberforce Road, Cambridge CB3 0WA, UK\label{aff147}
\and
Kavli Institute for Cosmology Cambridge, Madingley Road, Cambridge, CB3 0HA, UK\label{aff148}
\and
Department of Astrophysics, University of Zurich, Winterthurerstrasse 190, 8057 Zurich, Switzerland\label{aff149}
\and
Department of Physics, Centre for Extragalactic Astronomy, Durham University, South Road, Durham, DH1 3LE, UK\label{aff150}
\and
IRFU, CEA, Universit\'e Paris-Saclay 91191 Gif-sur-Yvette Cedex, France\label{aff151}
\and
Oskar Klein Centre for Cosmoparticle Physics, Department of Physics, Stockholm University, Stockholm, SE-106 91, Sweden\label{aff152}
\and
Univ. Grenoble Alpes, CNRS, Grenoble INP, LPSC-IN2P3, 53, Avenue des Martyrs, 38000, Grenoble, France\label{aff153}
\and
INAF-Osservatorio Astrofisico di Arcetri, Largo E. Fermi 5, 50125, Firenze, Italy\label{aff154}
\and
Dipartimento di Fisica, Sapienza Universit\`a di Roma, Piazzale Aldo Moro 2, 00185 Roma, Italy\label{aff155}
\and
Centro de Astrof\'{\i}sica da Universidade do Porto, Rua das Estrelas, 4150-762 Porto, Portugal\label{aff156}
\and
HE Space for European Space Agency (ESA), Camino bajo del Castillo, s/n, Urbanizacion Villafranca del Castillo, Villanueva de la Ca\~nada, 28692 Madrid, Spain\label{aff157}
\and
INAF - Osservatorio Astronomico d'Abruzzo, Via Maggini, 64100, Teramo, Italy\label{aff158}
\and
Theoretical astrophysics, Department of Physics and Astronomy, Uppsala University, Box 516, 751 37 Uppsala, Sweden\label{aff159}
\and
Institute for Astronomy, University of Hawaii, 2680 Woodlawn Drive, Honolulu, HI 96822, USA\label{aff160}
\and
Mathematical Institute, University of Leiden, Einsteinweg 55, 2333 CA Leiden, The Netherlands\label{aff161}
\and
School of Physics \& Astronomy, University of Southampton, Highfield Campus, Southampton SO17 1BJ, UK\label{aff162}
\and
Institute of Astronomy, University of Cambridge, Madingley Road, Cambridge CB3 0HA, UK\label{aff163}
\and
Department of Astrophysical Sciences, Peyton Hall, Princeton University, Princeton, NJ 08544, USA\label{aff164}
\and
Space physics and astronomy research unit, University of Oulu, Pentti Kaiteran katu 1, FI-90014 Oulu, Finland\label{aff165}
\and
Center for Computational Astrophysics, Flatiron Institute, 162 5th Avenue, 10010, New York, NY, USA\label{aff166}}    

%
%

\abstract{The MAMBO mock galaxy catalogue, based on the Millennium Simulation with empirically assigned galaxy properties, provides predictions of far-infrared (FIR) fluxes and physical parameters of \Euclid-detectable galaxies. We present the predicted FIR flux distributions and confirm that only the brightest \Euclid\ sources will be detectable in existing FIR surveys. To characterise the broader \Euclid population, we employ stacking to measure the mean dust properties as a function of stellar mass and redshift. We find dust temperatures and infrared luminosities increase with redshift across all mass bins, while dust masses remain roughly constant. FIR number counts from MAMBO show overall good agreement with observations, and the total infrared luminosity function (IRLF) reproduces published estimates across most redshift ranges, extending to $z \sim 10$. Comparing the Euclid Wide and Deep Surveys, we find that the EDS recovers the total IRLF to fainter luminosities and higher redshifts (up to $z \sim 6$ in $I_{\rm E}$), although its detectability falls below $80\%$ at $z > 4$, whereas the EWS becomes strongly incomplete beyond $z \sim 2$. We also examine the dependence of the IRLF on environment. Schechter fits indicate that the faint-end slope $\alpha$ flattens with redshift for cluster and protocluster galaxies, while remaining approximately constant for field populations. Imposing additional detection limits from \textit{Herschel}-PACS and SPIRE shows that only the most luminous ($L_{\rm IR} \gtrsim 10^{12.5} L_{\odot}$) galaxies remain detectable at $z \sim 4$, but the limited MAMBO area (3.14 deg$^2$) is inadequate for statistically robust ($>3\sigma$) constraints. Survey areas at least 30 times larger are required. Overall, the MAMBO FIR extension reproduces key number count and IRLF trends, provides realistic predictions for FIR-detected \Euclid\ galaxies, and highlights the importance of synergies with current (\textit{Herschel}, ALMA) and future FIR/sub-mm facilities (e.g. PRIMA, CCAT) to probe environmental dependence with sufficient depth and area.}

%
%
%
%

   \titlerunning{FIR predictions for \Euclid galaxy catalogues}
   \authorrunning{Euclid Collaboration: A. Parmar et al.}
\maketitle
%
%
%
%
\section{\label{sc:Intro}Introduction}

The luminosity and number density evolution of a sample of sources can usually be investigated using a luminosity function (LF, $\phi (L)\, \diff L$). This is simply defined as the comoving space density of galaxies per luminosity interval as a function of luminosity and redshift. It has been known since the launch of the first far-infrared (FIR) space telescopes that the study of the FIR and submillimetre (sub-mm) regime is vital for understanding galaxy formation and evolution \citep{Puget_1996A, Dole_2006, Devlin_2009}. Dust continuum emission, reprocessed from the UV emission of young stars, is the dominant contributor to the FIR luminosity of galaxies. Therefore, FIR luminosities are a probe for studying the cosmic star-formation history (CSFH). By measuring the infrared luminosity function (IRLF) at various redshifts, we can investigate the evolution of different galaxy populations. The IRLF can also be used to study specific galaxy types, for example, different galaxy morphologies, or galaxies in different environments. While the IRLF can be calculated at specific rest-frame wavelengths (e.g., \citealt{Gruppioni_2013, Gruppioni_2020}), it is conventionally defined as the integrated luminosity function for rest-frame IR wavelengths, $L_{\rm{IR}}$, in the range $8 \leq \lambda/\micron \leq 1000$.

Characterisation of the IRLF was made possible with the onset of FIR space telescopes, the Infrared Astronomical Satellite \citep[IRAS;][]{Neugebauer_1984_IRAS} and the Infrared Space Observatory \citep[ISO;][]{Kessler_1996}. With these instruments, the IRLF was explored for the first time \citep{Saunders_1990, Pozzi_2004, Serjeant_2001, Serjeant_2004}. However, these studies were limited to low redshifts ($z < 1$) due to sensitivity limits. It was not until the \Spitzer \citep{Werner_2004} that the IRLF at higher redshifts ($z > 1$) could be probed \citep{Floch_2005, Caputi_2007, Rodighiero_2010}. Since then, we have seen great advancement in FIR instruments, particularly the \Herschel-Space Observatory \citep{Pilbratt_2010}, which allowed us to probe the IRLF to $z > 4$, past the peak of cosmic star formation \citep{Gruppioni_2013, Hatsukade_2018, Wang_2019, Gruppioni_2020, Barrufet_2023, Fujimoto_2024}. These studies have focused mostly on the integrated IRLF or rest-frame monochromatic LF. 

\cite{Rieke_1986} and \cite{Isobe_1992} found that elliptical galaxies -- commonly located within clusters -- are on the whole less IR-luminous than spiral galaxies, which are much more rarely found in clusters. The optical LF of galaxies in different environments, such as field, groups, and clusters has been explored in several works which resulted in two different scenarios: one states that the LF depends on the environment, while the second is in favour of a universal LF. The advent of recent large-scale galaxy surveys has led to better determinations of the LF. Within the scenario that the LF is dependent on the environment, there is a broad consensus that the faint-end slope of the LF for field galaxies is fairly flat \citep{Loveday_1992, Marzke_1994, Lin_1996, Christlein_2003, dePropris_2003}, while other authors find that clusters of galaxies tend to exhibit a very steep faint-end slope \citep{Ferguson_1991, Valotto_1997}. \cite{Mo_2004} focused on the dependence of the LF on large-scale environment. They found that the characteristic luminosity, $L^{\rm{*}}$, increases with environment density. Conversely, the faint-end slope shows no such dependence on the environment density, unless the sample is sorted further into early- and late-type galaxies; late-type galaxies exhibit a flat faint-end, while the faint-end slope of early-type galaxies gets steeper with density. \cite{Zandivarez_2011} analysed the LF of galaxies in groups, finding that galaxies in high-density environments have brighter characteristic luminosities and steeper faint-end slopes than low-density environments. Studying the variation of the galaxy LF with group mass, they found that in high-density regions, galaxies in groups showed LF parameters that remained nearly unchanged across different group masses, in contrast to those in low-density regions.

Studies of the IRLF in (super)cluster environments have mostly been carried out with \textit{Spitzer} data. \cite{Bai_2006, Bai_2009} calculated the IRLF of the local ($z < 0.1$) clusters Coma and Abell $3266$, finding a similar bright-end shape for both clusters, and similar $L^*$ values between these clusters and other local field galaxies. \cite{Bai_2009} compared the IRLF of these two local clusters with two distant clusters ($z \sim 0.8$) and suggested a strong redshift evolution in both the $L^*$ values and the normalisation of the IRLF, with the more distant clusters consisting of more and brighter IR galaxies. 
\cite{Finn_2010} compared the IRLF of $16$ local ($z = 0.4$--$0.8$) galaxy clusters to a sample of coeval field galaxies and found that although the fraction of IR-luminous galaxies is significantly lower in clusters compared to the field, the $L_{\mathrm{IR}}$ of individual galaxies remains comparable. \cite{Biviano_2011} studied the evolution of the IRLF in different density regions of a supercluster. Within the Abell $1763$ supercluster located at $z = 0.23$, they defined three different regions: the cluster core; a large-scale filament; and the cluster outskirts (excluding the filament). They found that the filaments (intermediate-density region) contained the highest fraction of IR galaxies across all luminosities, whilst the cluster cores (high-density region) yielded the lowest fraction. The results of the cluster outskirts (low-density region) fell in-between the former two environments. \cite{Biviano_2011} also found that the number density of IR galaxies increased with redshift for all luminosities, in agreement with \cite{Bai_2009}. These various results of the (super)cluster IRLFs are limited to low redshift studies ($z < 1$) primarily due to the detection limits of the instruments used.

With the launch of the \Euclid Space Telescope \citep{Laureijs11, EuclidSkyOverview}, an era of refined optical and near-infrared (NIR) observations over a third of the entire sky is imminent.
The mission will include the Euclid Wide Survey \citep[EWS;][]{Scaramella-EP1} covering about \SI{1\,4000}{\deg\squared} over six years. This survey is estimated to observe down to a $5\sigma$ AB magnitude limit for a point-like source of $26.2$ in the visible band (\IE; \citealt{EuclidSkyVIS}) and an average of $24.4$ in the NIR bands (\YE, \JE, \HE; \citealt{EuclidSkyNISP, Schirmer-EP18}). A Euclid Deep Survey (EDS) will also be conducted over $53\ \rm{deg^2}$, probing two magnitudes deeper than the EWS. 

Multiple mock galaxy catalogues from different simulations were generated during the pre-launch phase to mimic \Euclid observations. One such mock catalogue is the MAMBO lightcone \citep[Mocks with Abundance Matching in BOlogna;][]{Girelli_2020} derived from the Empirical Galaxy Generator code \citep[\texttt{EGG};\footnote{\url{https://cschreib.github.io/egg/}}][]{Schreiber_2017}. The mock catalogue contains both passive and star-forming galaxies with simulated data including flux densities, luminosities, star-formation rates (SFRs) etc. in the \Euclid bands and various FIR bands. We aim to use the MAMBO lightcone data to predict FIR flux densities and physical properties of star-forming galaxies, and perform various calculations such as FIR number counts and IRLFs to compare with the literature. By doing this validation, we will test the reliability of the mock and determine whether it accurately represents the FIR-sub-mm regime. Furthermore, our goal is to study how dust-related properties, particularly $L_{\rm{IR}}$, vary with environment (i.e., whether a source is located in a cluster, protocluster, or field). With the mock extending to the FIR, we can study this regime, given the condition that galaxies are detectable by both \Euclid and FIR instruments, and how their properties change with environment. In this way, we can explore and provide estimates for the environmental dependence of the IRLF over a wider redshift range than that covered by previous observational studies. In particular, this is the first work to provide FIR predictions for \Euclid galaxy IRLFs.

This paper is structured as follows. In Sect.~\ref{sc:Mambo}, we describe how the MAMBO lightcone was generated to create the mock catalogue used in this work. Section~\ref{sc:FIR flux predictions} discusses the average FIR flux densities and corresponding average FIR properties of \textit{Eucild}-selected galaxies.
In Sect.~\ref{sc:Number counts} we present and discuss the FIR number counts using the MAMBO data. Various LFs are explored in Sect.~\ref{sc:LF}. Our conclusions are listed in Sect.~\ref{sc:Conclusion}.
In this paper, we adopt the same cosmology used to create the MAMBO mock to make the analysis consistent, unless specified otherwise. All magnitudes are given in the AB system.

\section{The MAMBO lightcone} \label{sc:Mambo}

MAMBO \citep{Girelli_2020} is a general procedure to simulate galaxy properties (stellar masses, SFRs, photometry, dust temperatures, metallicities, size, decompositions of bulge/disk, emission lines, and low-resolution spectra) of dark matter (DM) sub-haloes in lightcones, enabling the rapid creation of realistic multiwavelength mock data. In the present work, the MAMBO lightcone is generated from the Millennium DM $N$-body simulation \citep{Springel_2005}, rescaled to the 2013 \Planck cosmology\footnote{\Omm $ = 0.315$, \OmLa $ = 0.685$,  $\,h = 0.673$, $\,n_{\rm{s}}=0.961$, and $\sigma_8=0.826$ \citep{Planck_2014}.} and is based in particular on the lightcones created by \cite{Henriques_2015}, covering an area of \SI{3.14}{\deg\squared}. The lightcone used in the following work contains $7\,865\,440$ galaxies from $z\in[0.02,10]$ with (sub-)halo masses of $M_{\mathrm{sub}} > 1.7\expo{10}\,{M}_{\odot}\,h^{-1}$. MAMBO uses the sub-halo masses and their positions, and assigns galaxy properties following empirical prescriptions. In this way the halo properties (clustering, halo occupation distribution, and mass profile) are preserved, and the merger trees can be recovered from the original simulation, providing insights on the progenitors and descendants of galaxies and structures such as protoclusters.

To assign galaxy properties, the stellar mass is first derived through the stellar-to-halo mass relation \citep{Girelli_2020}, using \cite{Peng_2010} for the stellar mass function (SMF) at $z = 0$ from the Sloan Digital Sky Survey (SDSS), \cite{Ilbert_2013} for the SMF in COSMOS at $0.2 < z < 4$, and \cite{Grazian_2015} in CANDELS at $z \geq 4$. Each object is then assigned to a quiescent or star-forming galaxy type (with a random fraction of starburst galaxies) following a probability distribution derived from the ratio of the SMFs for the blue and red populations. As a result, galaxies with large stellar masses, which are more likely to be quiescent, are assigned to massive sub-haloes that predominantly inhabit the central regions of massive haloes, thus reproducing the relation between colour and environment.
All other galaxy properties are derived with a modified version of the public code \texttt{EGG} \citep{Schreiber_2017}. The resulting catalogues (excluding the FIR extension developed for this work) have been extensively validated \citep[see][for a thorough explanation]{Girelli_phd}. \texttt{EGG} relates stellar mass, redshift, and the galaxy type to other properties through observed scaling relations; the star-forming main sequence \citep{Schreiber_2015} is used to derive SFR, the fundamental plane relation \citep{Curti_2020} for metallicity, the Kennicutt--Schmidt law \citep{Kennicutt_1998} for \ha\, line luminosity, and distributions of sizes and line ratios from previous deep surveys. 

Rest-frame and observed photometry from ultraviolet (UV) to sub-mm are derived from the spectral energy distributions (SEDs) of the bulge and disc components. At optical and NIR wavelengths, the SEDs are derived from a pre-built library of templates from the \cite{Bruzual_2003} models, and the dust attenuated with the Calzetti law \citep{Calzetti_2000}, covering the $UVJ$ colours \citep{Williams_2009} of quiescent and star-forming galaxies. Infrared SEDs are derived from a set of libraries \citep{Chary_2001,Magdis_2012,Schreiber_2017} aimed at reproducing dust emission, characterised by the values of the infrared luminosity of dust and polycyclic aromatic hydrocarbons (PAHs) at $8 \leq \lambda/\micron \leq 1000$, the dust temperature, and the ratio of IR to $8\,\micron$ luminosity \citep[IR8;][]{Elbaz_2011} for dust and PAHs. The dust temperature and IR8 are derived as a function of redshift and stellar mass for star-forming and starburst galaxies. 

The lightcone derived with MAMBO and used in this work contains only galaxies, with no contribution from active galactic nuclei (AGN), while a version considering Type 1 and 2 AGN is included in \cite{Lopez_2024}. Potential implications or side effects of excluding this component will be discussed later. The lightcone also provides observed magnitudes and perturbed magnitudes that include expected measurement errors, such that the signal-to-noise ratio (S/N) corresponds to the expected limits for extended objects in the EWS. In this work we use the unperturbed magnitudes along with unperturbed redshifts, and AGN contribution is not included.

\section{Results}
\subsection{\label{sc:FIR flux predictions} MAMBO predictions for FIR galaxies}

\begin{figure*}
    \centering
    \includegraphics[scale=0.5]{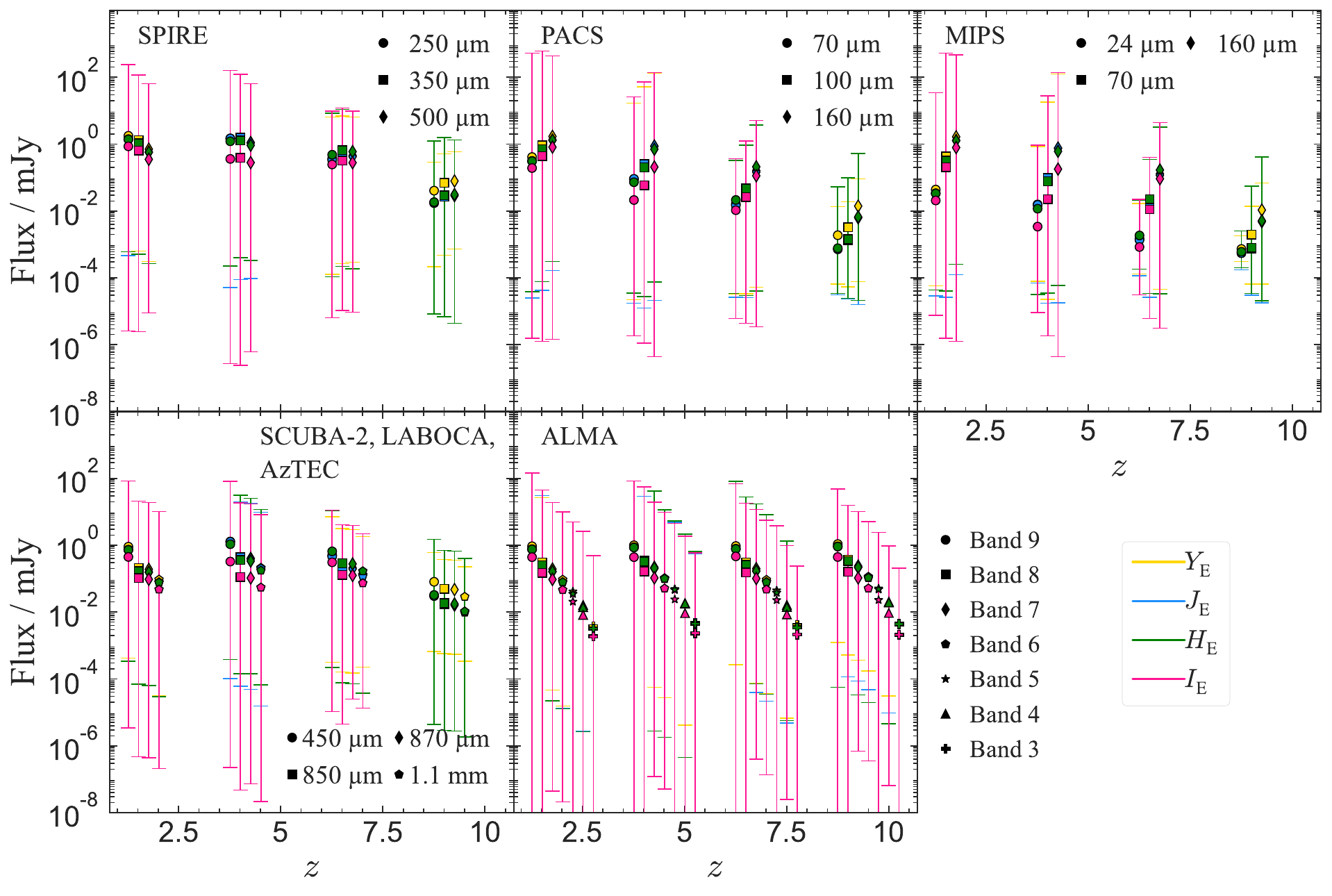}
    \caption{Predicted flux ranges of EWS-detectable MAMBO galaxies in various FIR/mm bands and redshift bins. Markers show the mean flux. The bands per subplot are offset in the x-axis for clarity.}
    \label{fig:FIR flux ranges}
\end{figure*}

To assess the detectability of \Euclid-detectable galaxies in the FIR--millimetre regime, we predict their flux distributions across multiple bands and redshift intervals. Figure~\ref{fig:FIR flux ranges} shows the predicted range of FIR/mm fluxes for EWS-detectable galaxies from MAMBO, separated by wavelength band and redshift. The predicted fluxes span several orders of magnitude, ranging from sub-microjansky levels to hundreds of millijanskys. Mean fluxes typically decrease with redshift, largely due to cosmological dimming and the increasing difficulty of detecting faint sources at early epochs. At $z > 7.5$, the \IE band yields no FIR/mm detections because it is redshifted shortward of the Lyman limit.

Across redshifts, for the MIPS and PACS bands, the mean flux increases with wavelength -- a reflection of SEDs peaking in the FIR and moving through bands due to redshifting. In contrast, the SPIRE bands ($250$, $350$, $500\,\micron$) show relatively flat mean fluxes with redshift, while for SCUBA-2, LABOCA, AzTEC, and ALMA ($450\,\micron$--$4\,$mm), mean fluxes decrease with wavelength because they sample the Rayleigh--Jeans tail of the dust SED. The lower bounds of predicted fluxes often drop below $10^{-5}\,$mJy, which reflects the saturation in source counts at the faint end since there is no noise to contend with.

Comparing these predicted flux ranges with the flux limits of the existing instruments highlights the challenges and opportunities for FIR/mm follow-up of \Euclid-detectable galaxies. For example, the \textit{Herschel}-PACS 3$\sigma$ confusion limits are approximately $5, 6,$ and $12.7\,$mJy at $70, 100,$ and $160\,\micron$, respectively \citep{Berta_2010}. Most of the predicted flux ranges in these bands at $z \sim 1$--$5$ fall below these thresholds, suggesting that only a fraction of the brightest star-forming PACS galaxies will be detectable. By $z \gtrsim 6$, the entire predicted flux ranges fall below these limits. Similarly, the SPIRE confusion limits (approximately $6$--$8\,$mJy; \citealt{Nguyen_2010}) intersect the predicted ranges just above the mean fluxes at intermediate redshifts, meaning that only the brightest \Euclid-detectable galaxies would be detected with SPIRE. For SCUBA-2, the $1\sigma$ depths from typical surveys (e.g., about $1\,$mJy at $850\,\micron$ in S2CLS; \citealt{Geach_2017}) exceed the majority of the $850\,\micron$ predicted fluxes between $z \sim 1$–$5$, and exceeds the entire predicted range at $z \gtrsim 7.5$. These comparisons highlight that, due to sensitivity and confusion limits, only a subset of \Euclid-detectable FIR emitters (usually the brightest) will be detected in archival \Herschel or SCUBA-2 datasets.

\begin{figure}
    \centering
    \includegraphics[width=\columnwidth]{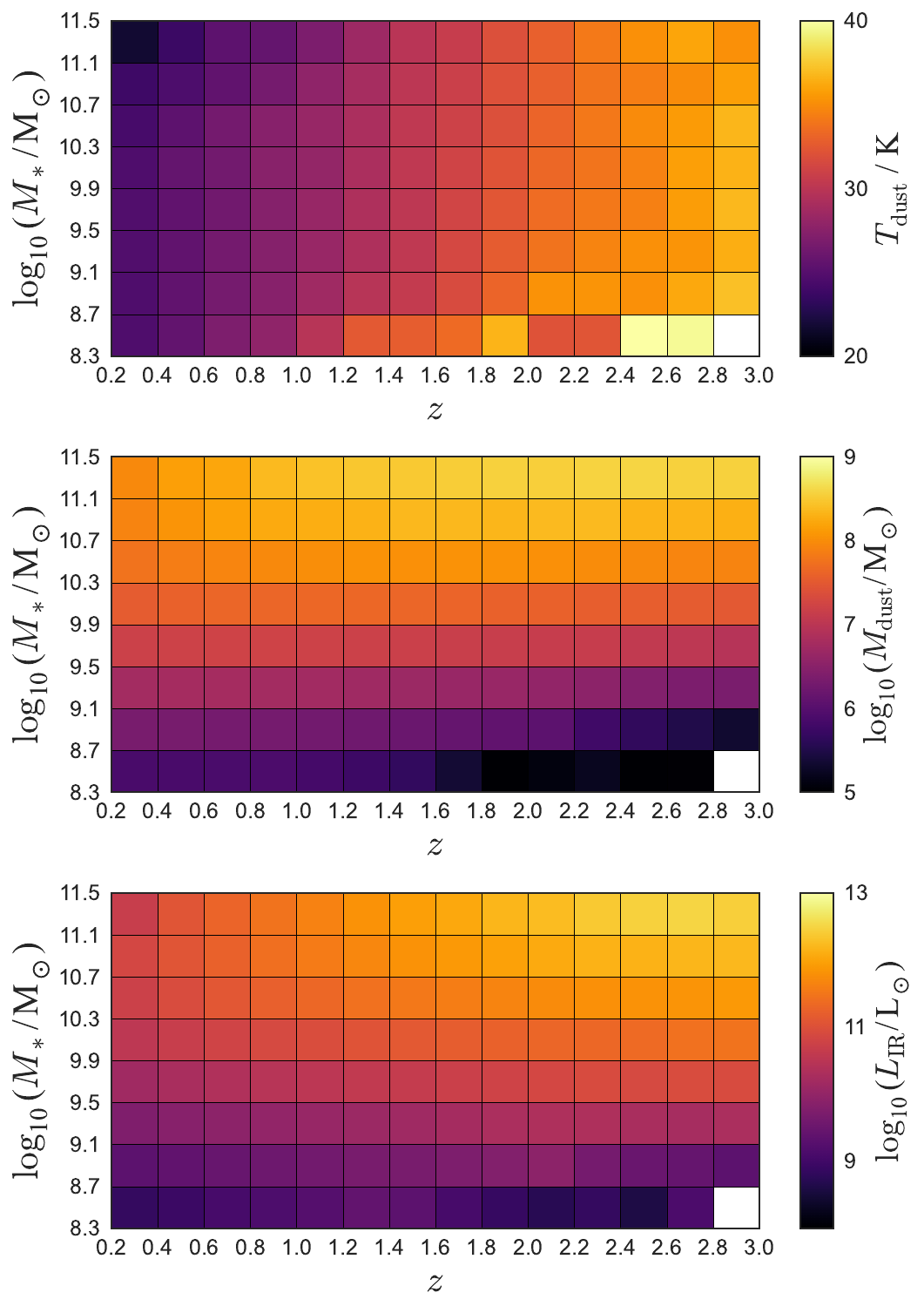}
    \caption{Average FIR physical properties of EWS-detectable galaxies as predicted by MAMBO. \emph{Top}: Average dust temperatures ($T_{\text{dust}}$). \emph{Middle}: Average dust masses ($M_{\text{dust}}$). \emph{Bottom}: Average IR luminosities ($L_{\text{IR}}$).}
    \label{fig: colourmaps}
\end{figure}

Although the fainter sources will not be detected individually, their collective signal can still be measured through stacking analyses. By statistically combining the FIR/mm flux densities of binned \Euclid-detectable galaxies, stacking enables the recovery of average FIR fluxes well below nominal detection thresholds, thereby providing insights into the dust-obscured star formation and infrared properties of the broader \Euclid population. Below, we explore how the mean FIR emission ($100$--$850\,\micron$) varies with redshift and stellar mass for the \Euclid mock galaxies. This analysis can be used as a comparison with ongoing and future studies of \Euclid-selected star-forming galaxies that are detected in existing FIR surveys. We bin the EWS-detectable galaxies by redshift (from $z = 0.2$–$3$ in steps of $0.2$) and by stellar mass (from $8.3 < \log_{10}(\mathit{M}_{*} / M_{\odot}) < 11.5$ in steps of $\log_{10}(\mathit{M}_{*} / M_{\odot}) = 0.4$ dex). We choose these bins to match a similar analysis in \cite{Q1-SP071} using existing \Euclid and FIR data. The average FIR fluxes for each bin are reported in Table~\ref{tab:average fluxes}.

Within each stellar mass and redshift bin, we extract the average dust temperature $T_{\text{dust}}$, dust mass $M_{\text{dust}}$, and IR luminosity $L_{\text{IR}}$ from the MAMBO catalogue. These quantities are visualised in Fig.~\ref{fig: colourmaps}. There are no EWS-detectable galaxies in any of the FIR bands at $z = 2.9$ for a stellar mass of $\log_{10}(\mathit{M}_{*} / M_{\odot}) = 8.5$. We find a clear increase in $T_{\text{dust}}$ with redshift, rising from approximately $20$\,K at $z \sim 0.2$ to $40$\,K at $z \sim 3$. We find no strong trend in $T_{\text{dust}}$ with stellar mass at fixed redshift. Similarly, $L_{\text{IR}}$ increases by more than three orders of magnitude across the redshift range. This implies enhanced dust-obscured star formation at earlier times, particularly in the most massive systems, and agrees with the increased $T_{\text{dust}}$ at high redshifts. The $M_{\text{dust}}$ remains relatively constant with redshift, with more massive galaxies displaying systematically larger dust masses.

The combination of increasing dust temperature and infrared luminosity is consistent with what is observed in high-redshift galaxies. Their elevated specific SFRs mean that intense UV radiation from young, massive stars heats the surrounding dust more strongly, raising the dust temperature. This enhanced heating both increases the total infrared luminosity -- as more stellar light is reprocessed by dust -- and shifts the peak of the dust SED toward shorter wavelengths.

Using the stacked FIR flux densities from MAMBO and applying the same SED-fitting procedure as in \cite{Q1-SP071}, we find that the resulting SEDs are consistent with those derived from real \Euclid\ data \citep{Q1-SP071}. The mean physical FIR properties extracted from these fits -- comparable to the results in Fig.~\ref{fig: colourmaps} -- are also consistent with the initial parameterisations adopted to simulate the lightcone. In particular, the mean dust temperatures follow the relation given in Equation~(14) of \cite{Schreiber_2017}, while converting the mean luminosities into mean SFRs yields results consistent with Equation~(10) of \cite{Schreiber_2017}.

\subsection{\label{sc:Number counts} FIR number counts}

\begin{figure*}[h]
    \includegraphics[width=2\columnwidth]{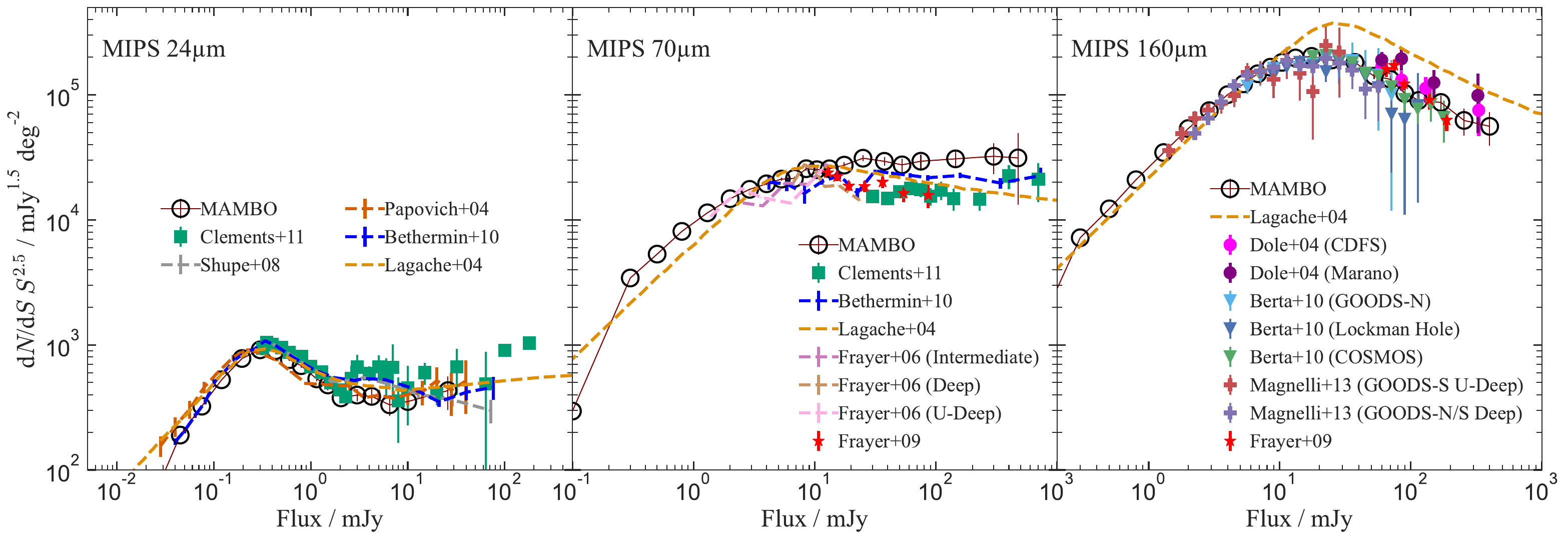}
    \caption{Differential number counts of the MAMBO mock for the \textit{Spitzer}-MIPS bands. Data and models from the following works are shown for comparison: \cite{Lagache_2004}, \cite{Dole_2004}, \cite{Papovich_2004}, \cite{Frayer_2006}, \cite{Frayer_2009}, \cite{Shupe_2008}, \cite{Berta_2010}, \cite{Clements_2011}, \cite{Bethermin_2010_spitzer}, and \cite{Magnelli_2013}.}
    \label{spitzer counts}
\end{figure*}

\begin{figure*}[h]
    \includegraphics[width=2\columnwidth]{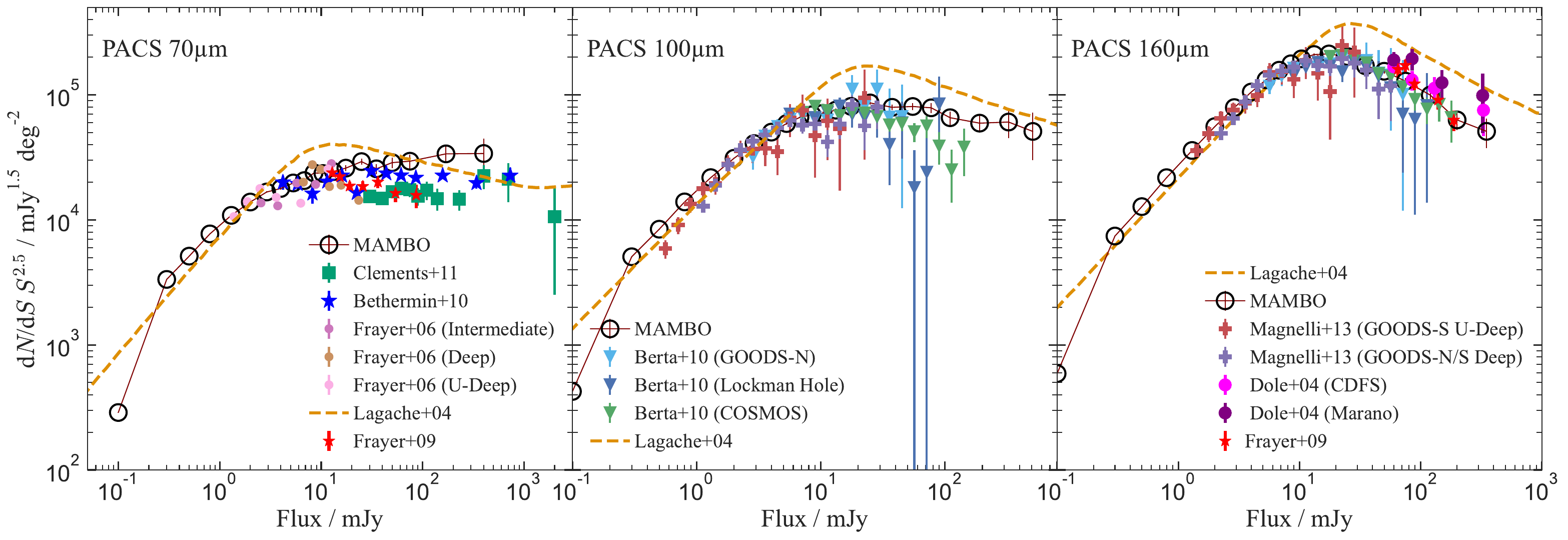}
    \caption{Differential number counts of the MAMBO mock for the \textit{Herschel}-PACS bands. 
    The legend in the $100\,\micron$ panel also applies to the $160\,\micron$ panel, and vice versa. 
    Data and models from the following works are shown for comparison: \cite{Lagache_2004}, \cite{Frayer_2006}, \cite{Berta_2010}, \cite{Clements_2011}, \cite{Bethermin_2010_BLAST}, and \cite{Magnelli_2013}.}
    \label{pacs counts}
\end{figure*}

\begin{figure*}[h]
    \includegraphics[width=2\columnwidth]{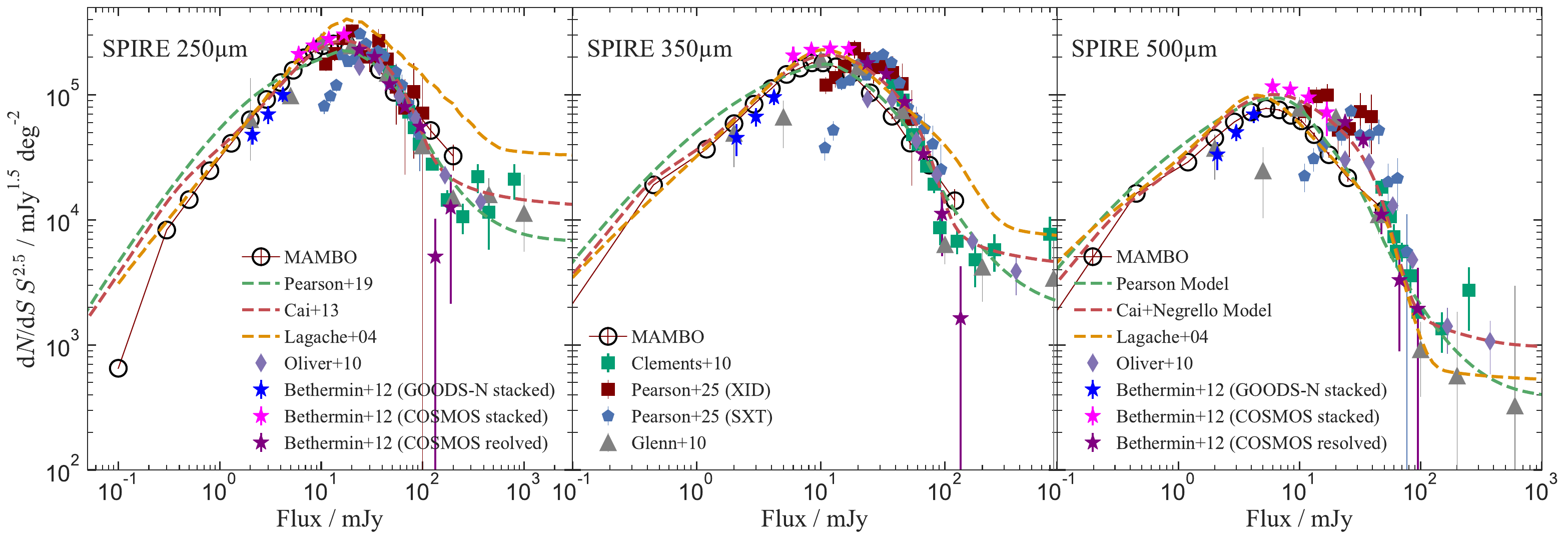}
    \caption{Differential number counts of the MAMBO mock for the \textit{Herschel}-SPIRE bands. 
    The legend in the $250\, \micron$ and $350\, \micron$ panels apply to all three panels. 
    Data and models from the following works are shown for comparison: \cite{Lagache_2004}, \cite{Frayer_2006}, \cite{Berta_2010}, \cite{Clements_2011}, \cite{Bethermin_2010_BLAST}, and \cite{Magnelli_2013}.}
    \label{spire counts}
\end{figure*}

\begin{figure*}[h]
    \includegraphics[width=2\columnwidth]{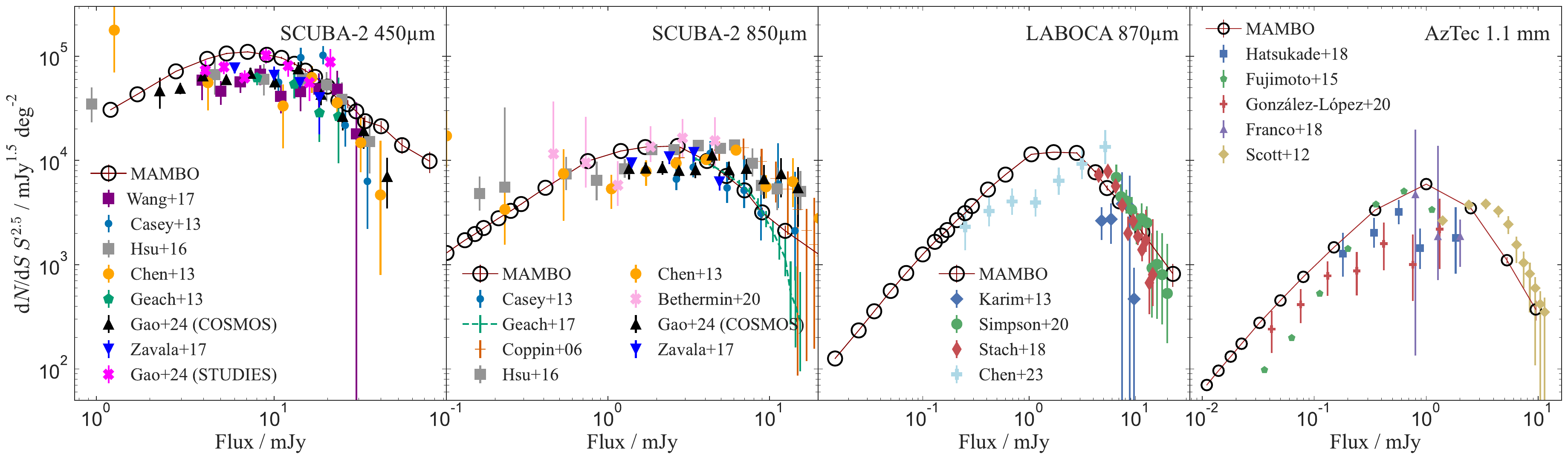}
    \caption{Differential number counts of the MAMBO mock for the SCUBA-2 bands, Apex-LABOCA $870\, \micron$ band, and AzTec $1.1\,$mm band. Data and models from the following works are shown for comparison: \cite{Wang_2017}, \cite{Casey_2013}, \cite{Geach_2013}, 
    \cite{Gao_2024}, \cite{Zavala_2017}, \cite{Lagache_2004}, \cite{Coppin_2006}, \cite{Knudsen_2008}, \cite{Zemcov_2010}, \cite{Borys_2003}, \cite{Bethermin_2020}, \cite{Karim_2013}, \cite{Simpson_2015}, \cite{Chen_2023},
    \cite{Hsu_2016}, \cite{Simpson_2020}, and \cite{Stach_2018}.}
    \label{scuba laboca counts}
\end{figure*}

\begin{figure*}[h]
     \centering
\includegraphics[width=2\columnwidth]{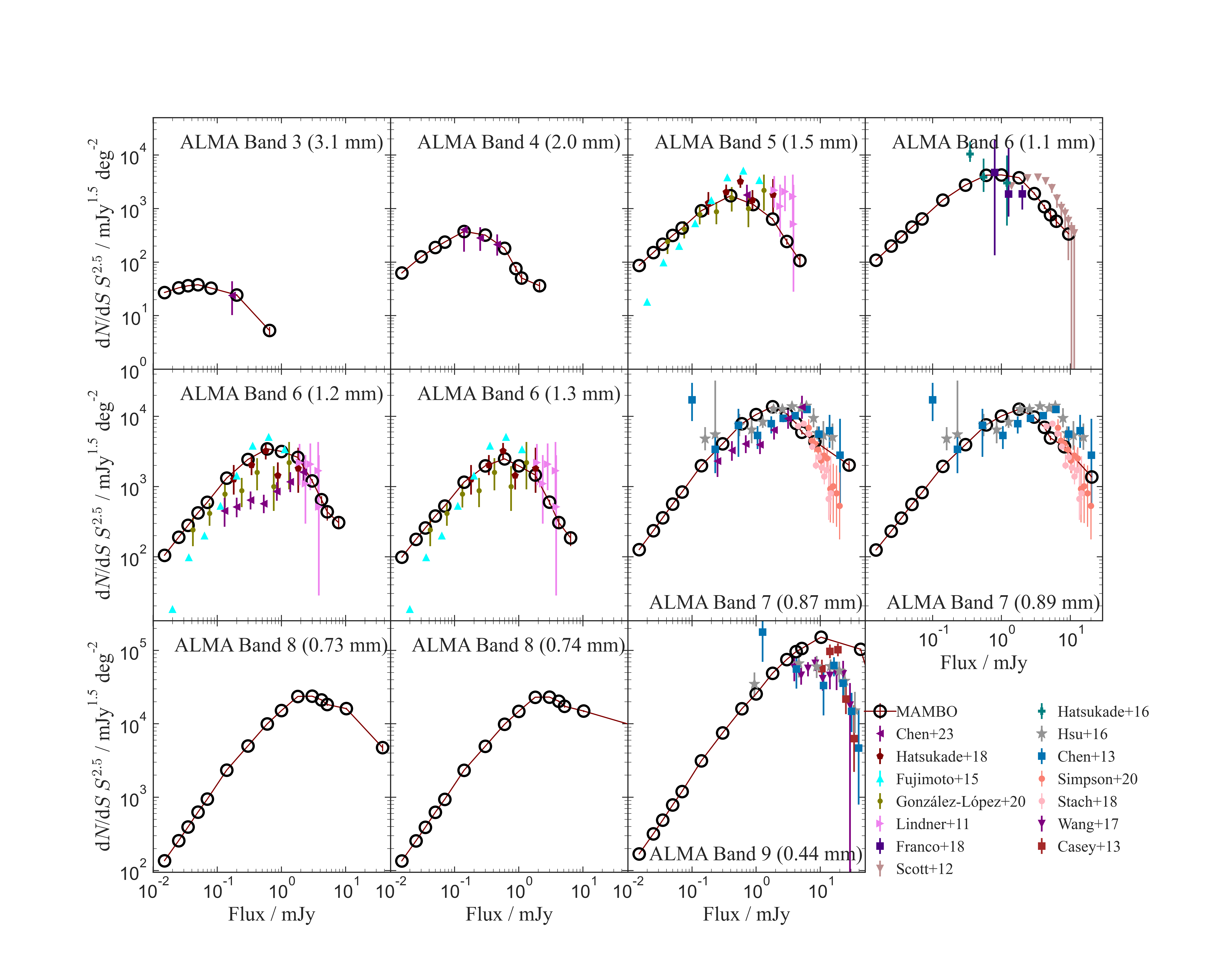}
    \caption{Differential number counts of the MAMBO mock for various ALMA bands. Data and models from the various works are shown for comparison.}
    \label{alma diff counts}
\end{figure*}

In this section, we examine the mock's ability to reproduce observed FIR number counts. We calculate the surface density of sources per flux density interval, $\diff N/\diff S$, for all galaxies in the mock. These counts are a simple measure of a population's abundance and an important tool for model comparison. Figures~\ref{spitzer counts}--\ref{alma diff counts} show the Euclidianised\footnote{The differential counts $\diff N/\diff S$ are multiplied by $S^{2.5}$ so a non-evolving Euclidean universe appears flat, making any deviations clearer to identify.} differential counts ($\diff N/\diff S\,S^{2.5}$) for the MAMBO mock at various FIR wavelengths, compared with the published estimates. The cumulative counts are presented in Appendix~\ref{int counts}.

In Fig.~\ref{spitzer counts}, the \textit{Spitzer}-MIPS number counts from MAMBO at $24\, \micron$ and $160\, \micron$ are generally well matched to the literature, with the exception of the \citet{Lagache_2004} model in the $160\, \micron$ differential counts, whose predicted peak deviates from the plotted data.
There is tension between MAMBO and the data for \textit{Spitzer}-MIPS (Fig.~\ref{spitzer counts}) and \textit{Herschel}-PACS (Fig.~\ref{pacs counts}) at $70\, \micron$. For both these $70\, \micron$ bands, we see an overestimation in the MAMBO counts at fluxes greater than approximately $10\,$mJy. We calculate a maximum discrepancy at the bright-flux end of the differential counts of $5 \sigma$ and $6\sigma$ for MIPS and PACS, respectively. We define $\sigma$ based on the uncertainty of the published counts that show the largest discrepancy from the MAMBO counts. This discrepancy at $70\, \micron$ could partly be a result of the dust temperature used in the MAMBO mock. The $70\, \micron$ emission, at all redshifts, is on the Wien tail of an SED for the majority of dust temperatures. This means that the $70\, \micron$ emission is susceptible to contributions from hot dust. The mock could include hot dust which may result in an overestimation of the number counts at the bright, and generally low-$z$ end of $70\, \micron$, but has less effect on the FIR wavelengths. While AGN typically dominate mid-IR emission at shorter wavelengths (e.g., the $24\,\micron$ band), their contribution at $70\,\micron$ is generally modest. Therefore, neglecting AGN emission is unlikely to significantly affect the number counts at $70\,\micron$, especially at the bright end, where star formation dominates the dust heating. In addition to the discrepancy at MIPS $70\, \micron$, the bright-flux end of the MAMBO $100\, \micron$ (Fig.~\ref{pacs counts}) differential counts are in excess of the published estimates (by about $2$--$3\,\sigma$) at fluxes greater than $40\,$mJy. The \cite{Lagache_2004} model here is also in disagreement with the differential counts peak of the other data.

For the \textit{Herschel}-SPIRE counts (Fig.~\ref{spire counts}), the $250\, \micron$ counts are generally in agreement with the data. However, in the differential counts at $350\, \micron$ and $500\, \micron$, we see MAMBO underestimating the number counts compared to the existing data, at flux densities between approximately $10$ and $50$\,mJy. At $350\, \micron$, the \cite{Pearson_2019} model also underestimates the observed differential counts and at $500\, \micron$, both the \cite{Pearson_2019} and \cite{Lagache_2004} models underestimate the observed differential counts. This disagreement between models and observation is explored in \cite{Bethermin_2017}. Their own simulation also underestimates the \textit{Herschel}-SPIRE counts at similar fluxes to MAMBO ($5$--$50$\,mJy). In limited angular resolution instruments such as \textit{Herschel}-SPIRE, one reason for this could be the blending of several sources within the SPIRE beam. \cite{Bethermin_2017} highlight the importance of taking into account resolution and clustering effects before comparing simulated catalogues with observations. They also find that their number counts (at $5$--$50$\,mJy) extracted from simulated maps are in better agreement with the observed data. Creating and analysing simulated maps from the MAMBO lightcone is not something explored in this work, but would be a path for future work.

A discrepancy is also observed for SCUBA-2 at $450\, \micron$ (Fig.~\ref{scuba laboca counts}) where the mock overpredicts the differential number counts at about $7\,$mJy, with the largest discrepancy being about $4\sigma$. The only difference between the MAMBO data for the two SCUBA-2 wavelengths is the type of transmission curve used in the simulation. For $850\, \micron$ the transmission curve measured during the instrument's performance was used and so it includes the effects of atmospheric absorption. For $450\, \micron$, the model transmission curve\footnote{\url{https://www.eaobservatory.org/jcmt/instrumentation/continuum/scuba-2/filters/}} was used and thus did not include the effects of atmospheric absorption. A $450\, \micron$ transmission curve including the effects of atmospheric effects was not available. Since the atmosphere is significantly more opaque at $450\, \micron$, the use of the model transmission curve likely results in an overestimate of the effective throughput and, consequently, the observed fluxes in the mock catalogue. This may help explain the overprediction in number counts at $450\, \micron$ around $7\,$mJy, a discrepancy that is not as apparent at $850\, \micron$, where realistic transmission losses were properly included. The SCUBA-2 $850\, \micron$ MAMBO counts show an overall good agreement with the literature. The MAMBO $870\, \micron$ number counts (Fig.~\ref{scuba laboca counts}) are generally in good agreement with the published data at the bright-flux end, but MAMBO seems to overpredict the counts at intermediate fluxes, with a maximum discrepancy of about $6\sigma$. However, this discrepancy arises from comparing the mock counts with only one available dataset. 
A similar overprediction is seen for the AzTEC $1.1\,$mm counts (Fig.~\ref{scuba laboca counts}) where MAMBO portrays an excess at about $1\,$mJy, with a maximum discrepancy of $3 \sigma$ from the differential counts. 

The ALMA counts (Fig.~\ref{alma diff counts}) show an overall good agreement with results from the literature, although some discrepancies are evident at the bright-flux end for ALMA Band~$7$. Since no Band~$7$ number counts were available for direct comparison, we adopt the $850\, \micron$ counts as a proxy, given the similar central wavelengths. While the mock and observed counts are broadly consistent at fainter fluxes, the MAMBO mock predicts a higher number density of bright sources. This discrepancy may in part reflect the idealised nature of the mock which assumes perfect source extraction and no confusion or blending; these effects may bias observational counts at the bright end. Additionally, minor differences in the effective wavelengths ($870/890\, \micron$ versus $850\, \micron$) may introduce SED-dependent flux offsets that further contribute to the apparent excess in the MAMBO counts. In Fig.~\ref{alma diff counts}, for ALMA Band~$9$ counts, MAMBO predicts a higher abundance of sources at the peak and bright-flux end of the differential counts than the published results. We see a maximum discrepancy of $7 \sigma$ at this bright-flux end.

There are some wavelengths at which the bright-end tail of the literature counts extend past the bright-end tail of the MAMBO number counts. This is seen at the SPIRE wavelengths as well as MIPS $24\, \micron$ and PACS $70\, \micron$. The truncation in the MAMBO counts is due to the relatively small area of the MAMBO lightcone.

Overall, the MAMBO mock is generally able to reproduce the observed abundance of galaxies at the various wavelengths for both integral and differential counts.

\subsection{\label{sc:LF} IRLF}

\subsubsection{\label{sc:IR LF} The total IR luminosity function}

We derive the total IRLF for all galaxies in the MAMBO lightcone, applying an $L_{\mathrm{IR}}$ cut of $\logten(L_{\mathrm{IR}}/L_\odot) > 8$. This luminosity threshold ensures inclusion of sources that are luminous in the FIR. Our LFs are calculated using the $1/V_{\mathrm{max}}$ method \citep{Schmidt_1968}. The MAMBO sample is sorted into redshift bins over the redshift range of the lightcone, $z\in[0.02,10]$, as in \cite{Gruppioni_2013} and \cite{Rodighiero_2010}. The sources are sorted into $L_{\mathrm{IR}}$ bins, over the range $8 \leq \logten(L_{\mathrm{IR}}/L_\odot) \leq 14$, each with a width of $0.5$ dex. In each $L_{\mathrm{IR}}$ bin we compute the comoving volume available to each source within that bin, defined as $V_{\mathrm{max}} = V_{z_{\mathrm{max}}} - V_{z_{\mathrm{min}}}$. Since we include all galaxies within the mock for this case, we define $V_{z_{\mathrm{max}}}$ and $V_{z_{\mathrm{min}}}$ as the upper and lower boundaries, respectively, of the redshift bin under consideration. The IRLF is then calculated using
\begin{equation} \label{LF eq}
    \phi(L,z) = \frac{1}{\Delta \logten(L_{\rm{IR}}/L_\odot)} \sum_{i} \frac{1}{V_{\mathrm{max},i}},
\end{equation} 
where $\Delta \logten(L_{\rm{IR}}/L_\odot)$ is the width of the logarithmic luminosity bin and $V_{\mathrm{max},i}$ is the comoving volume over which the $i$th galaxy can be observed.

\begin{figure*}[ht]
\includegraphics[angle=0,width=2\columnwidth]{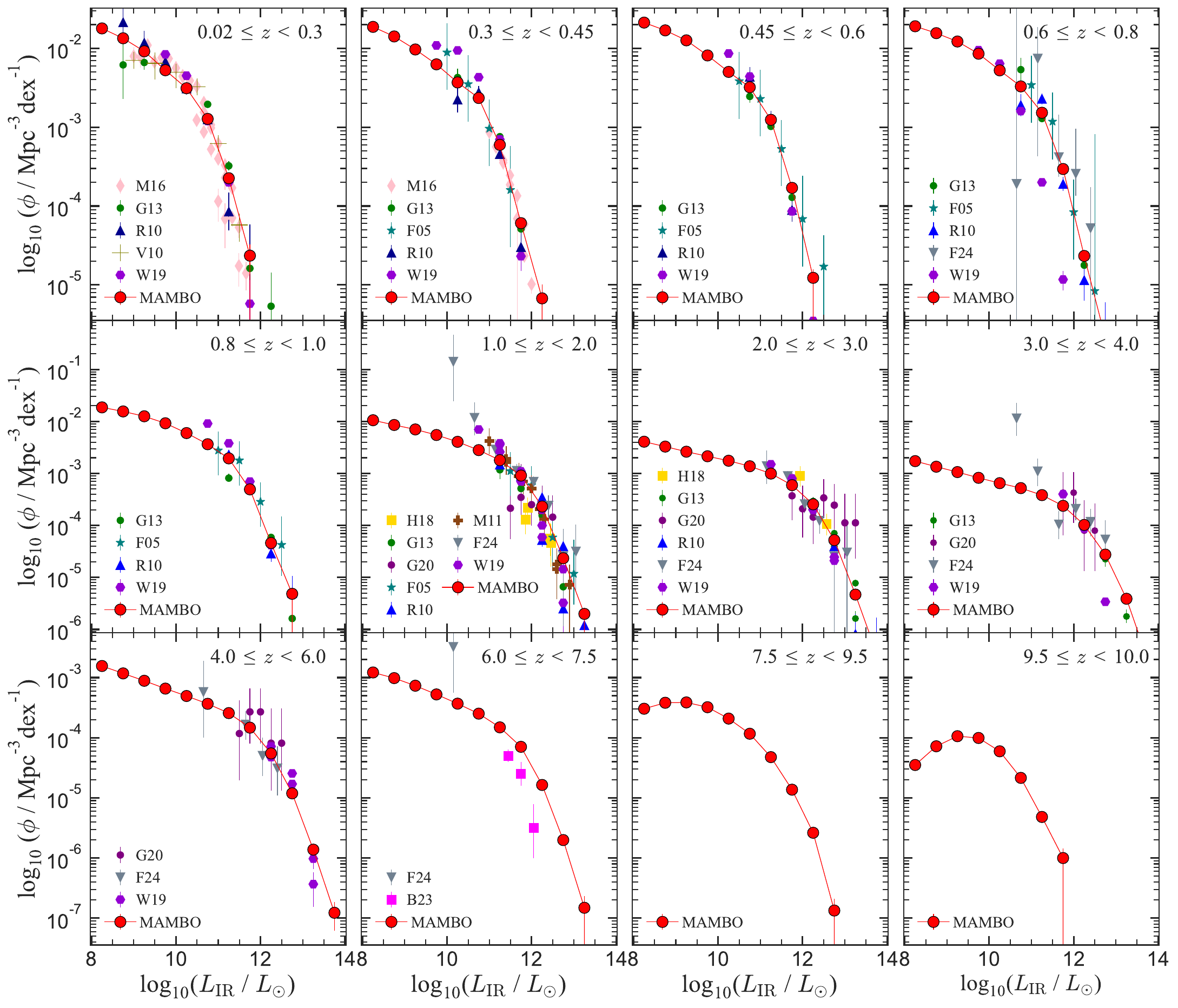}  
\caption{Total IRLF of all the galaxies within the \textit{\Euclid} MAMBO mock, shown as the red circles. Various data that best match the defined redshift bins are plotted for comparison. Papers referenced in short-form in the plot are defined as follows: G13: \cite{Gruppioni_2013}, M16: \cite{Marchetti_2016}, R10: \cite{Rodighiero_2010}, V10: \cite{Vaccari_2010}, F05: \cite{Floch_2005}, M11: \cite{Magnelli_2011}, F24: \cite{Fujimoto_2024}, G20: \cite{Gruppioni_2020}, B23: \cite{Barrufet_2023}, W19: \cite{Wang_2019}, and H18: \cite{Hatsukade_2018}.}
\label{LF all}
\end{figure*}

We show the total IRLF in Fig.~\ref{LF all}. The MAMBO mock is denoted by the red circles and relevant literature results are over-plotted.
We find excellent overall agreement with the literature. At \mbox{$z > 6$}, the scarcity of observational data renders comparison more difficult. Some notable cases of discrepancy are discussed below.
At $z = 1.0$--$2.0$ and $z = 3.0$--$4.0$, the faint end of the \cite{Fujimoto_2024} IRLF has a steeper slope than the MAMBO IRLF, with a difference of $2 \sigma$. Note that it is the only published data to extend to such faint luminosities. 
In the $z = 6.0$--$7.5$ redshift bin, the MAMBO IRLF exhibits a slightly higher normalisation than the \cite{Barrufet_2023} results, with a maximum discrepancy of $4\sigma$. The \cite{Barrufet_2023} data are based on a UV-selected sample, which implies that their sample does not account for extremely dust-obscured sources that are UV-faint, leading to the lower normalisation of the \cite{Barrufet_2023} LF compared to the MAMBO IRLF. The \cite{Fujimoto_2024} result for this redshift bin is also inconsistent with the MAMBO IRLF ($2\sigma$ discrepancy). \cite{Fujimoto_2024} consistently predict a higher density of sources at the faint end of the luminosity function compared with MAMBO. The faint-end slope of the IRLF is difficult to constrain, as reflected in the sparsity of data at higher redshifts. However, in their analysis, \cite{Fujimoto_2024} note that while their faint-end probes $1$--$2\,$ dex deeper than other published results, their results overall agree within $1\sigma$. Most of these comparison data cover luminosity ranges that focus on $L^*$, while \cite{Fujimoto_2024} probe to fainter luminosities ($\sim10^{10} L_{\odot}$) beyond the knee of the IRLF. Their faint-end results are consistent with MAMBO within $2\sigma$ uncertainties. Additionally, their sample consists of sources with ALMA data, providing greater sensitivity and higher resolution which can allow exploration below the single-dish confusion limit and reduce the impact of the incompleteness of identifying faint sources. 
We also cannot rule out the possibility that, since their results consistently predict a steeper slope at the faint end than MAMBO and the available data, MAMBO may be underestimating the density of fainter sources. Overall, however, the MAMBO mock reproduces the majority of observational IRLFs and, in some cases, probes to fainter luminosities than the existing data.

\subsubsection{\label{sc:euclid flux lim LF} IRLF; EWS versus EDS}

\begin{figure*}[ht]
    \centering
    \includegraphics[angle=0,width=2\columnwidth]{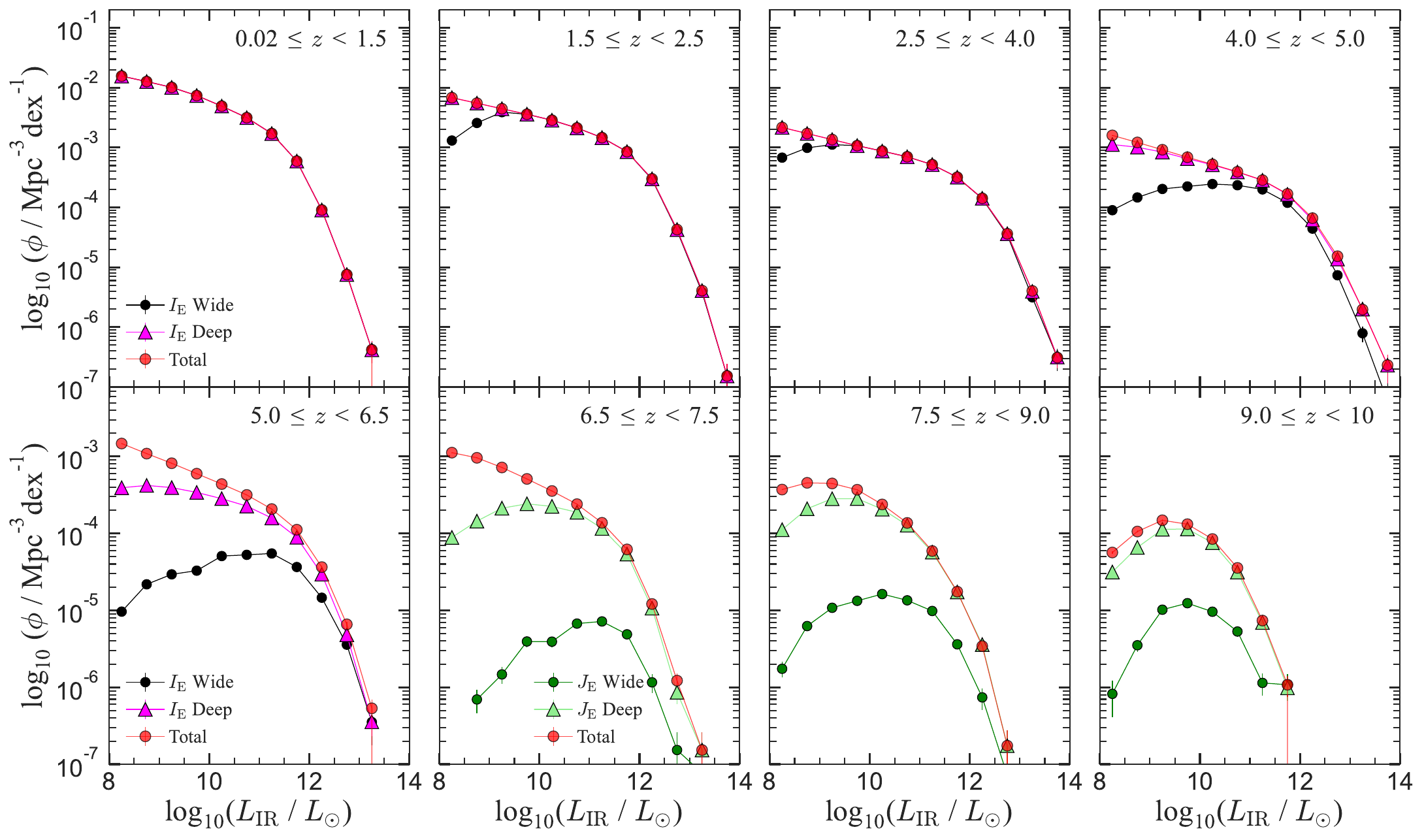}
    \caption{$1/V_{\rm max}$-derived IRLF for the EDS-detectable sample compared to the EWS-detectable sample. Additionally plotted for comparison is the total IRLF of all the galaxies within the \textit{Euclid} MAMBO mock, shown by the red circles.}
    \label{LF fir flux lim deep}
\end{figure*}

To explore the differences between the EWS and EDS, we compare the IRLF of all galaxies in the sample after applying AB magnitude cuts corresponding to the $5\sigma$ point-source detection limits for each \Euclid band (\YE, \JE, \HE, and \IE) in the EWS and EDS, respectively, as listed in Table~\ref{Euclid mag limits}. For this comparison, we extract sources detectable in each band given the respective survey’s magnitude limits and apply the same $L_{\mathrm{IR}}$ cut as before.
The comoving volume $V_{\mathrm{max}}$ is computed for each source. In this case, $z_{\mathrm{min}}$ corresponds to the lower boundary of the redshift bin, while $z_{\mathrm{max}}$ is defined as the lesser of the upper redshift boundary and the redshift at which the source falls below the survey’s detection limit in the relevant \Euclid band. The resulting IRLF is then calculated using Eq.~\eqref{LF eq} and shown in Fig.~\ref{LF fir flux lim deep}. For a clearer comparison, we only plot the $\IE$ band for the majority of our redshift bins. The total IRLF (red circles) represents the full IRLF and any deviations from this are attributed to incompleteness introduced by the \Euclid magnitude cuts. In other words, there are some IR galaxies that are undetectable by \Euclid at any redshift.

\begin{figure}[h]
    \centering
    \includegraphics[angle=0,width=1\columnwidth]{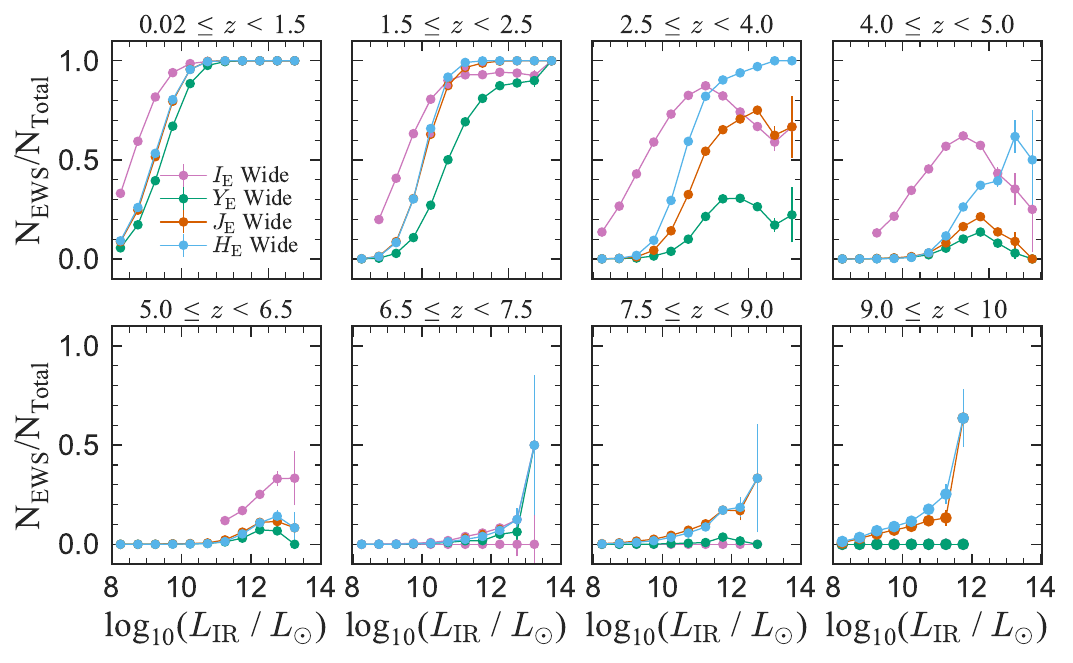}
    \caption{Fraction of detectable IR sources in the different \Euclid bands in various $z$ bins for the EWS.}
    \label{LF fir flux lim EWS completeness}
\end{figure}

\begin{figure}[h]
    \centering
    \includegraphics[angle=0,width=1\columnwidth]{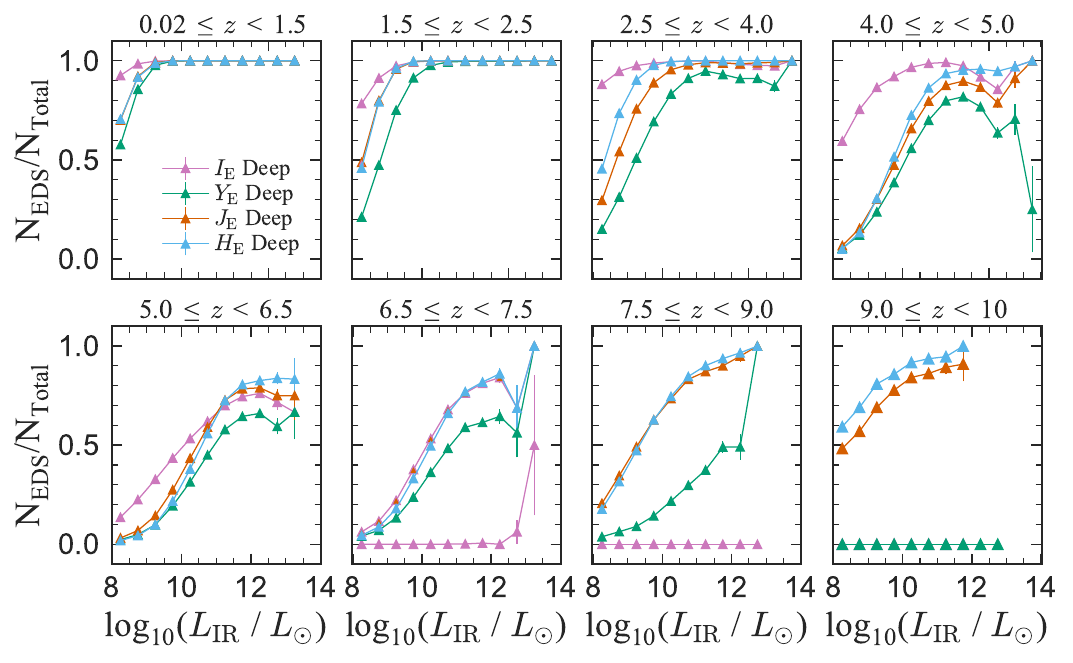}
    \caption{Fraction of detectable IR sources in the different \Euclid bands in various $z$ bins for the EDS.}
    \label{LF fir flux lim EDS completeness}
\end{figure}

We calculate the ratio of \Euclid-detectable IR sources to the total population of IR sources for each $(L,z)$ bin. These results are shown in Figs.~\ref{LF fir flux lim EWS completeness} and \ref{LF fir flux lim EDS completeness} for the EWS and EDS, respectively. In the first two redshift bins of Fig.~\ref{LF fir flux lim EWS completeness}, the fraction of IR-detectable galaxies in the \IE Wide sample is greater than $80\%$ at approximately $\logten(L_{\rm{IR}}/L_\odot) > 10$, after which the percentage falls below $80\%$ in the remaining redshift bins.
In the first three redshift bins of Fig.~\ref{LF fir flux lim EDS completeness}, $100\%$ of all IR galaxies are detectable by the EDS at $\logten(L_{\rm{IR}}/L_\odot) > 9.5$ leading to the overlap between the EDS and total IRLF in Fig.~\ref{LF fir flux lim deep} at these luminosities.
Beyond $z = 4$, the fraction of the EDS-detectable sample falls with redshift. The downturn in the \IE band for the EWS at high $L_{\rm IR}$ in the $2.5 < z < 4$ and $4 < z < 5$ bins likely reflects the increasing fraction of heavily dust-obscured, optically redder and UV-fainter systems at the highest infrared luminosities such that a larger fraction falls below the EWS \IE detection threshold. This effect is much less pronounced in the EDS fields, whose greater depth recovers a larger fraction of these optically faint, high-$L_{\rm IR}$ galaxies, supporting an interpretation driven primarily by depth-limited detectability rather than intrinsic changes in the IR population.

At \mbox{$z = 7.5$}, no optical light is detected in the \IE band, and thus no IRLF is shown in Fig.~\ref{LF fir flux lim deep} for this band. For the highest two redshift bins, we instead compare the IRLF of the \JE band for the EWS and EDS. Although MAMBO predicts that galaxies will be visible at these redshifts, the EDS \JE sample only detects $80\%$ or more of the total sample at approximately $\logten(L_{\rm{IR}}/L_\odot) > 11$, whereas, for the same redshifts the EWS \JE band mostly recovers less than $30\%$ of the total IR sample. Finally, we test whether the choice of luminosity bin size affects the shape of the IRLF by recalculating it with smaller and larger bin widths. No significant changes in the overall IRLF shape are found across the tested binning schemes.

\begin{table}[h]
    \centering
      \caption{$5\sigma$ point-like source magnitude limit of each \Euclid band in the EWS and EDS, from \cite{Scaramella-EP1} and \cite{EP-Selwood}, respectively.}
    \begin{tabular}{c c c c}
       \hline
       \hline
       \noalign{\vskip 1.5pt}
       Band & $\lambda_{\mathrm{eff}}$ [\AA] & EWS limiting mag & EDS limiting mag \\
       \noalign{\vskip 1.5pt}
       \hline
       \noalign{\vskip 1pt}
       \IE & $7200$ & $26.2$ & $28.2$ \\
       \YE & $10\,810$ & $24.3$ & $26.3$ \\
       \JE & $13\,670$ & $24.5$ & $26.5$ \\
       \HE & $17\,710$ & $24.4$ & $26.4$ \\
       \hline
    \end{tabular}
    \label{Euclid mag limits}
\end{table}

\subsubsection{\label{sc:Environment LF} IRLF in different environments}

We also examine how the IRLF depends on environment. To do this, we sort galaxies into three categories based on their environment; i.e., whether they are located within a cluster, protocluster, or field, based on the definitions outlined below.

We first apply a lower redshift cut of $z = 1.5$ to the MAMBO mock. This threshold ensures that the redshift range includes many protoclusters, since large-scale structures are not yet fully assembled at these epochs. Galaxies are further selected using the magnitude criteria $\HE \leq 24.25$ or $\IE \leq 25.10$, corresponding to the expected $5\sigma$ detection limits of the EWS for extended sources.

Protocluster members are defined as galaxies associated with the same collapsed cluster by $z = 0$, where the final halo mass must satisfy $M \geq 10^{14}\,M_\odot$. This lower mass threshold provides an operative definition for identifying massive protoclusters. In contrast, cluster members are defined as galaxies that reside within the same halo at the redshift they are observed, where the halo mass must be $M \geq 10^{13.25}\,M_\odot$. This threshold aligns with the mass limit used in the \Euclid Cluster Finder Challenge \citep{Adam_2019}. Cluster member galaxies are available only in the redshift range \mbox{$1.5 \leq z \leq 4.0$} because of the original constraints of the GAEA lightcone \citep{Hirschmann_2016}; therefore, no cluster galaxies are included at $z > 4$ for consistency between lightcones.
There is an overlap between galaxies classified as cluster and protocluster members. This is expected, since a galaxy may reside in a cluster at a given redshift while also being part of a larger halo system that will evolve into a massive cluster by $z = 0$. Consequently, the environment-dependent luminosity functions derived here are not mutually exclusive.

Field galaxies are defined to inhabit regions of average or low density. That being said, there could be cases of galaxies found outside clusters (i.e., in the field) that are located in regions with densities larger than the average. At the same time, galaxies on the outskirts of protoclusters, and hence characterised as protocluster member galaxies, are likely to be in average density environments. Therefore, if all cluster and protocluster member galaxies were removed from the original MAMBO mock, the remaining galaxies: (1) might include some located in high-density regions; and (2) might not constitute a complete sample, since some galaxies would have been excluded by the protocluster definition. While this approach introduces these potential limitations, it provides a practical definition suitable for comparative analysis and so we choose to define field galaxies as those that remain after the cluster and protocluster members are removed.

\begin{figure*}[h]
    \includegraphics[width=2\columnwidth]{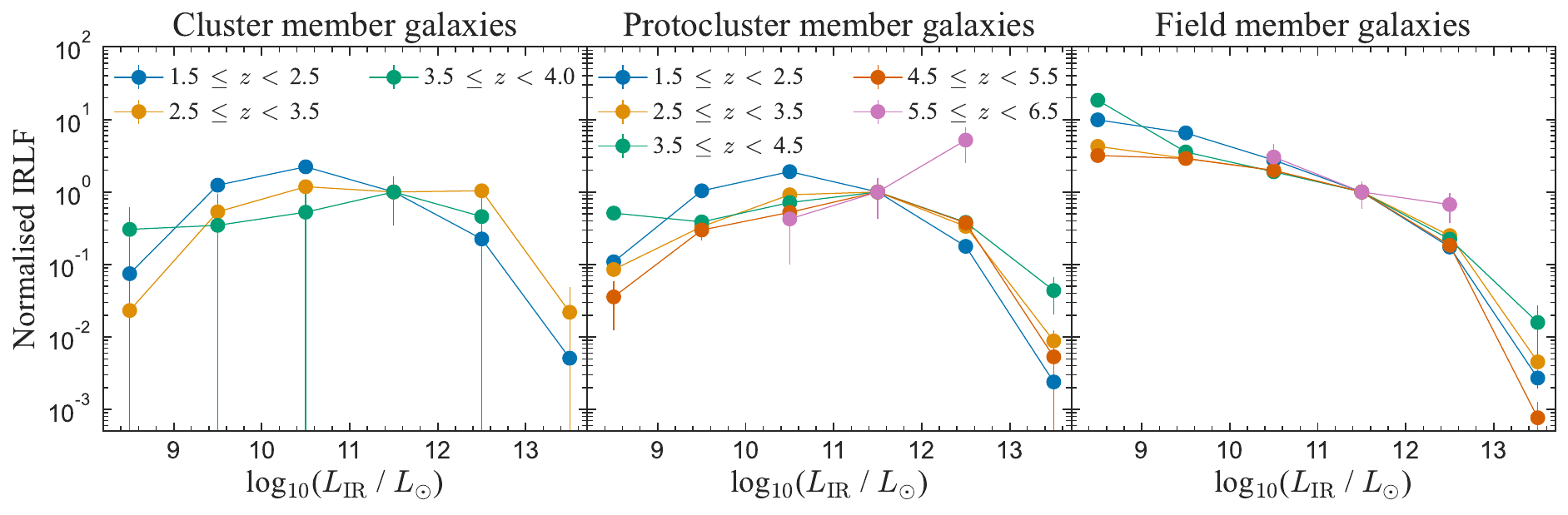}
    \caption{Completeness-corrected IRLF of MAMBO mock galaxies based on their environment: cluster, protocluster, or field. The LFs are normalised such that $\logten(\phi / \rm{Mpc}^{-3} dex^{-1}) = 1$ at $\logten(L_{\mathrm{IR}} / L_\odot) = 11.5$ for each redshift bin. The legend for protocluster member galaxies also applies to field member galaxies.}
    \label{LF env comparison normalised}
\end{figure*}

In Fig.~\ref{LF env comparison normalised}, we present the completeness-corrected IRLF of EWS-detectable MAMBO galaxies classified by environment for a range of redshift intervals. Since the comoving volume occupied by each environment per redshift bin is not accounted for, we do not compare the normalisation of the IRLFs directly. Instead, we compare their shapes and how they evolve with redshift. To facilitate this comparison, we normalise each IRLF such that $\logten(\phi / \rm{Mpc^{-3} \rm{dex^{-1}}}) = 1$ at $\logten(L_{\mathrm{IR}}/L_\odot) = 11.5$ within each redshift bin.

\begin{figure}[h]
    \includegraphics[width=\columnwidth]{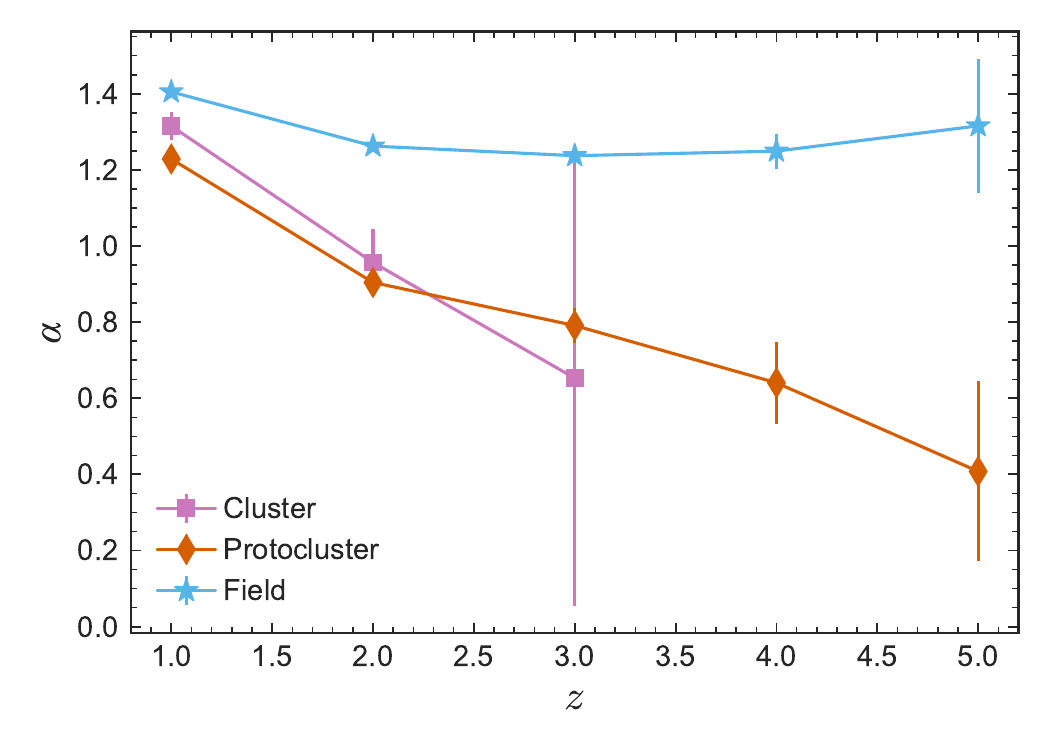}
    \caption{Redshift evolution of the completeness-corrected faint-end slope parameter $\alpha$ for different environments.}
    \label{alpha comparison bright}
\end{figure}

We restrict our Schechter function fits to the IRLFs in Fig.~\ref{LF env comparison normalised} to the range $10.5 \leq \logten(L_{\mathrm{IR}}/L_\odot) \leq 13.5$, since this is where the EDS and EWS IRLFs are more complete. This choice ensures that the fits are less biased by incompleteness and flux limits, which can artificially flatten or steepen $\alpha$ when low-luminosity bins are not robustly constrained. Although this excludes the classical faint end of the IRLF, it yields a more robust fit to the well-sampled portion of the data. The resulting $\alpha$ values should therefore be interpreted as effective slopes across this brighter luminosity range rather than as true faint-end slopes. As such, the absolute values of $\alpha$ are less robust than the relative trends with redshift and environment. Our results are shown in Fig.~\ref{alpha comparison bright}.

For cluster and protocluster environments, we observe a flattening of the $\alpha$ parameter with increasing redshift. For field galaxies, the $\alpha$ parameter remains approximately constant with redshift. The observational analysis of (super)cluster IRLFs is limited to low redshifts ($z < 1$), which makes it difficult to compare with the MAMBO predictions which probe $z \geq 1.5$. For field galaxies, there is a greater amount of published data available. There is broad consensus that the luminosity of IR galaxies increases while their number density decreases with redshift \citep{Caputi_2007, Magnelli_2011, Gruppioni_2013, Gruppioni_2020, Traina_2024}, however, we cannot compare the evolution of $L^{*}$ and $\phi^{*}$ from our Schechter function fits with Fig.~\ref{LF env comparison normalised} since the functions are not scaled to the comoving volume of each environment.

\cite{Bai_2009}, who studied the IRLF of local clusters ($z \leq 0.06$), fixed the faint-end slope parameter to $\alpha = 1.41$ in their Schechter function fits for galaxies with $\log_{10}(L_{\mathrm{IR}} / L_{\odot}) \gtrsim 9$. \cite{Bai_2006} examined the IRLF of the Coma cluster ($z < 0.1$) and also explored its dependence on local density. They found a shallower faint-end slope ($\alpha \sim 1.0$), again fitting for $\log_{10}(L_{\mathrm{IR}} / L_{\odot}) \gtrsim 9$, and observed a deficit of IR-luminous galaxies in the dense cluster core relative to lower-density outskirts. At $z = 0.23$, \cite{Biviano_2011} fitted a single power law to the IRLF of the A1763 cluster and found a faint-end slope of $\alpha \sim -2.0$, applying a lower luminosity cut of $\log_{10}(L_{\mathrm{IR}} / L_{\odot}) \sim 10.4$. We note that in this power-law parameterisation, $\alpha$ is negative by convention, with steeper slopes (larger absolute values) indicating a more rapid increase in number density toward lower luminosities, which is physically expected for most luminosity functions. This increase in number density is also evident, for the same luminosity range as in \cite{Biviano_2011}, in Fig.~\ref{LF env comparison normalised} across most redshift bins.

To further assess the observational reach of the EWS, we apply an additional detection constraint that requires galaxies to be observed both by \Euclid and by a NIR/FIR instrument, specifically, \textit{Herschel}-PACS \citep{Poglitsch_2009} and \Herschel-SPIRE \citep{Griffin_2010}. A SPIRE detection is defined as a greater than $3\sigma$ detection in each band, with flux density thresholds based on confusion noise: $17.4\,$mJy at \SI{250}{\micron}, $18.9\,$mJy at \SI{350}{\micron}, and $20.4\,$mJy at \SI{500}{\micron} \citep{Nguyen_2010}. For PACS, we adopt the $3\sigma$ flux density limits from the \Herschel GOODS-S survey, with \Herschel GOODS-S probing a little deeper than \Herschel GOODS-N: $0.8\,$mJy at \SI{100}{\micron} and $2.4\,$mJy at \SI{160}{\micron} \citep{Elbaz_2011}.

\begin{figure}[h]
    \includegraphics[width=\columnwidth]{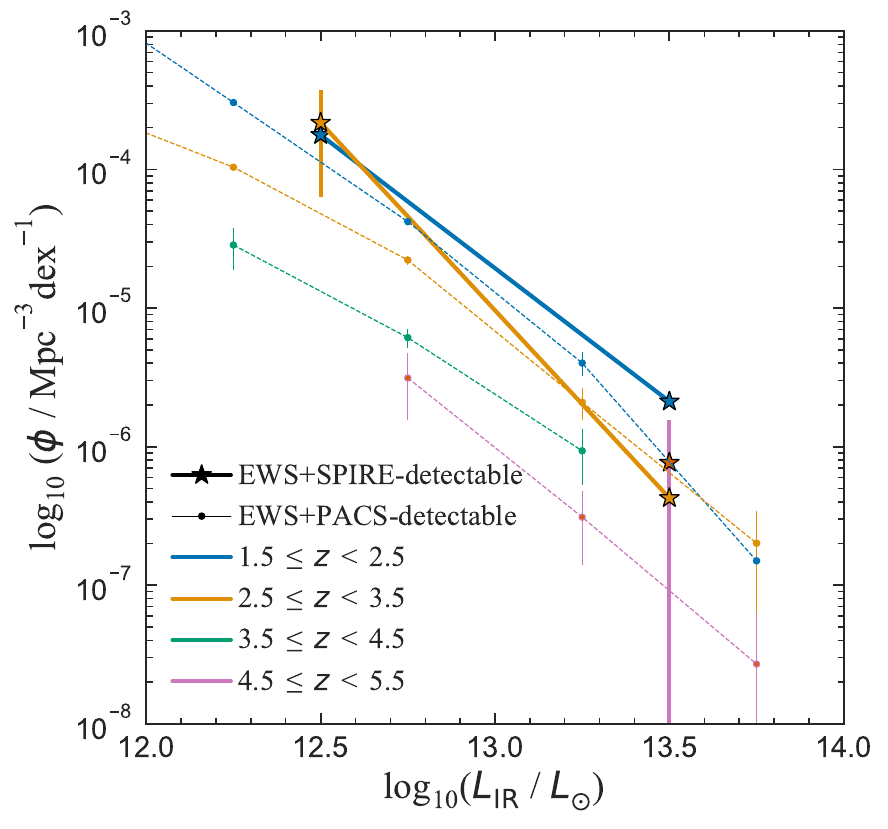}
    \caption{Completeness-corrected IRLF of all MAMBO galaxies that are potentially detected by the EWS and SPIRE/PACS. 
    Very few galaxies remain after the EWS$+$SPIRE limits are applied, so we do not see much evolution in the IRLF. For the EWS$+$PACS limits, more galaxies are detectable and we see a redshift evolution of the IRLF.}
    \label{LF MAMBO comparison}
\end{figure}

Prior to applying these limits to the environment-dependent IRLFs, we compute the IRLF for all MAMBO galaxies potentially detected by both EWS and PACS/SPIRE. We first filter the lightcone with the EWS detection limits and then further select only those sources that meet the PACS/SPIRE thresholds. The IRLFs are then recalculated using these filtered samples. In Fig.~\ref{LF MAMBO comparison}, we show the completeness-corrected IRLF for the EWS$+$SPIRE-detectable sample and the EWS$+$PACS-detectable sample. 
The IRLFs for the EWS$+$PACS-detectable galaxies display greater variation than those that are EWS$+$SPIRE-detectable due to Poisson statistics, because of the increased depth of the \Herschel GOODS-S survey and the correspondingly larger galaxy sample. As such, we see that the small number of galaxies being selected based on the SPIRE flux limits hinders the effectiveness of this IRLF. Galaxies are present only over a small luminosity range, yielding limited interpretive value. With the deeper limits of the PACS survey, a greater sample of galaxies remain after the selection limits are applied. This results in a clear redshift evolution of the IRLF for the EWS$+$PACS-detectable sample. 

\begin{figure*}[h]
\centering
\includegraphics[width=1.5\columnwidth]{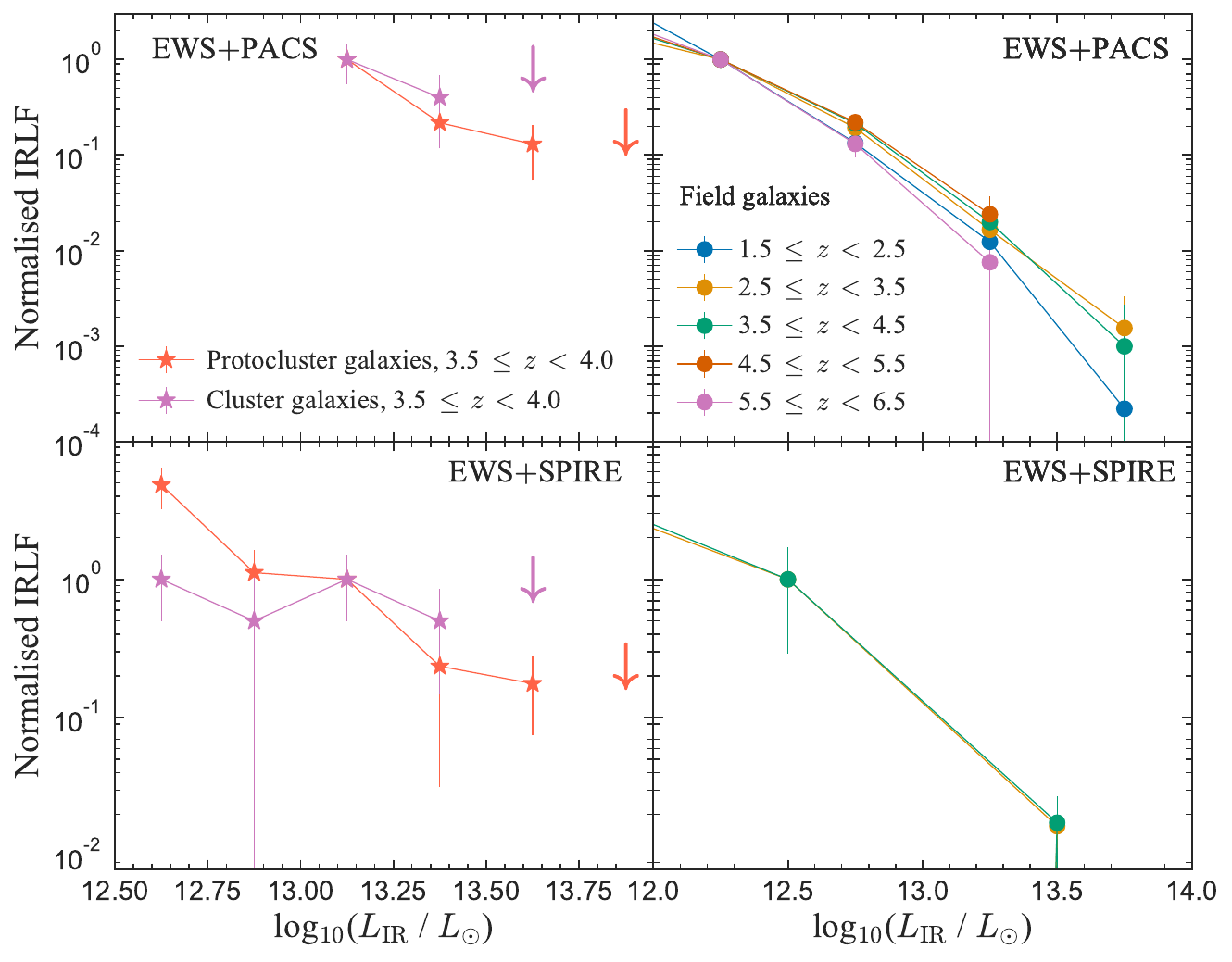}
\caption{\emph{Top row}: Completeness-corrected IRLF of MAMBO mock galaxies based on their environment given that they are detected by the EWS and \textit{Spitzer}-PACS. The cluster and protocluster LFs are normalised at $\logten(L_{\mathrm{IR}}/L_\odot) = 13.125$ and the field LFs at $12.25$. $3\sigma$ upper limits, denoted by a downwards arrow, are shown where there is only one source per luminosity bin. The legend in the top row also applies to the bottom row.
\emph{Bottom row}: Completeness-corrected IRLF of MAMBO mock galaxies based on their environment, given that they are detected by the EWS and \Herschel-SPIRE. The cluster and protocluster apparent LFs are normalised at $\logten(L_{\mathrm{IR}}/L_\odot) = 13.125$ and the field LFs at $12.5$.}
\label{LF fir env comparison normalised}
\end{figure*}

With the addition of the SPIRE and PACS limits,
we then calculate the completeness-corrected IRLF for the environmental-dependent sample and show our results in Fig.~\ref{LF fir env comparison normalised}. In the top row, the EWS$+$PACS limits are applied. Here, the IRLF for cluster and protocluster members has been normalised at $\logten(L_{\mathrm{IR}}/L_{\odot}) = 13.125$ and the field IRLFs at $\logten(L_{\mathrm{IR}}/L_{\odot})= 12.25$. 
In the bottom row, the EWS$+$SPIRE limits are applied. Here, the IRLF for cluster and protocluster members are also normalised at $\logten(L_{\mathrm{IR}}/L_{\odot}) = 13.125$ and the field IRLFs at $\logten(L_{\mathrm{IR}}/L_{\odot}) = 12.5$. In both the top and bottom rows, any luminosity bin with only one source is shown as a $3\sigma$ upper limit denoted by a downward arrow.

Cluster member galaxies are present only until $z = 4$ and it is in this high-redshift bin ($z=3.5$--$4$) where the EWS$+$SPIRE is able to detect them, but only over a small range at the highest luminosities; $12.6 \leq \log(L_{\mathrm{IR}}/L_\odot) \leq 13.6$. At lower redshifts, cluster member galaxies are predicted to be undetectable simultaneously by the EWS and SPIRE. Similarly, protocluster members are detectable only in the $z=3.5$--$4$ redshift bin. 
The shape of the IRLF for field galaxies at different redshifts remains unchanged over the small luminosity range shown. It should be noted that, due to the small area of the MAMBO lightcone (approximately \SI{3}{\deg\squared}), only a small number of galaxies remain after the EWS$+$SPIRE detection limits are applied, leading to the large Poisson errors seen in Fig.~\ref{LF fir env comparison normalised}. The EWS$+$PACS limits (top row of Fig.~\ref{LF fir env comparison normalised}) give similar results for cluster and protocluster members as the bottom row with EWS$+$SPIRE limits. The only galaxies visible are those in the $z=3.5$--$4$ redshift bin. The IRLF for EWS$+$PACS-detectable field galaxies is comparable to the IRLF in Fig.~\ref{LF env comparison normalised}. We see a clear decreasing trend for all redshifts in the IRLF for luminosities larger than $10^{11.75}\,L_\odot$, with little redshift variation in the shape of the function. 

However, given that the PACS observations of the \textit{Herschel} GOODS-S field cover $100\,$arcmin$^{2}$ in about $200$ hours, and PACS observations of a similar area and depth to MAMBO do not exist, these results are only a modest theoretical insight into the IRLF for field galaxies, and the reality of achieving this is unlikely without a new FIR mission.

The environment dependence seen in Fig.~\ref{LF env comparison normalised} for the high-luminosity end ($L_{\mathrm{IR}} > 10^{12}\, L_\odot$) suggests that this is a relation worth studying further, but it is clear that this will require a much larger survey area, for both simulation and observational data. Given our results for Fig.~\ref{LF fir env comparison normalised}, specifically for the EWS$+$SPIRE-detectable sample, we estimate that a mock with an area at least $30$ times greater than the current MAMBO mock is required to achieve IRLF results that are statistically reliable (i.e., greater than $3\sigma$). Existing large-scale \textit{Herschel} surveys such as HerMES \citep{Oliver_2012} and H-ATLAS \citep{Valiante_2016, Bourne_2016} have much larger survey areas (\SI{380}{\deg\squared} and \SI{600}{\deg\squared}, respectively) than the MAMBO lightcone. Performing a similar analysis with these surveys will provide a substantial number of FIR- and \Euclid-detectable sources for a statistically useful study on the environmental dependency of the IRLF using existing data, albeit we are limited by the instruments' sensitivity. With \textit{Herschel} no longer in operation, attention must shift to other existing facilities -- such as ALMA and NOEMA -- which observe in the millimetre regime and can conduct wide surveys that sample the same dust-emission SEDs, albeit at longer wavelengths. Furthermore, future instruments such as the Cerro Chajnantor Atacama Telescope (CCAT; \citealt{Aravena_2022}), operating between $0.3$–$3\,\mathrm{mm}$, will be capable of covering large survey areas with high sensitivity, enabling improved constraints on the Rayleigh–Jeans tail and peak of the FIR SED. The PRobe far-Infrared Mission for Astrophysics (PRIMA; \citealt{Moullet_2023}) is a future telescope concept, which, if selected, will provide additional large-area surveys at wavelengths that overlap with \textit{Herschel} but probe to deeper sensitivities.

\section{Conclusions} \label{sc:Conclusion}

With the FIR extension of the MAMBO lightcone, we investigated the predicted FIR properties of \textit{Euclid}-detectable galaxies and validated MAMBO's capacity to reproduce observed number counts and IRLFs. Furthermore, in preparation for upcoming \Euclid observations, we investigated its capability to detect NIR/FIR-detectable sources and assess the dependence of IRLFs on the galaxy environment. The main results of this study are summarised as follows.

\begin{enumerate}
    \item The predicted FIR flux densities of star-forming galaxies follows the expected trends with increasing redshift and highlights that only the brightest FIR galaxies will be detectable by \Euclid. Their predicted dust temperatures and IR luminosities generally increase with redshift, as expected, and their dust masses remain relatively constant with redshift.
    
    \item The MAMBO lightcone reproduces FIR number counts well across most bands. In some cases, the predicted number counts extend to fainter fluxes than the existing counts, offering model predictions at the faint end to compare with future data.

    \item The total IRLF calculated for the entire MAMBO sample shows excellent agreement with the published data across most redshift bins (see Fig.~\ref{LF all}). At redshifts \mbox{$z > 6$}, the sparsity of observational data prevents strong conclusions.

    \item Comparing the IRLF for the EWS- and EDS-detectable samples (Fig.~\ref{LF fir flux lim deep}), we find similar IRLF shapes at $0.02 < z < 4$, particularly at $\logten(L_{\mathrm{IR}}/L_\odot) > 10$ where the fraction of IR-detectable sources is the highest ($> 80\%$). With increasing redshift, we see a greater deviation from the total IRLF by the EWS sample, while the greater depth of the EDS is able to select a higher fraction of IR galaxies out to $z = 6$.
    At \mbox{$z > 6$}, the \IE band is no longer effective due to the redshifted Lyman limit, although galaxies remain detectable in the NIR bands with both surveys, up to the highest redshifts albeit at much smaller fractions. Our ratio of IR-detectable sources in the \Euclid surveys will enable observed \Euclid IRLFs to be corrected for the missing populations entirely undetectable by \Euclid.

    \item By fitting the Schechter function to the environment-dependent IRLF, we observe a moderate flattening of the faint-end slope $\alpha$ with increasing redshift, for cluster and protocluster environments. The slope for field galaxies remains approximately constant with redshift.

    \item By applying additional NIR/FIR detection constraints based on the flux limits of \Herschel-PACS and SPIRE, we explore the IRLF's environmental dependence (Fig.~\ref{LF fir env comparison normalised}). Cluster and protocluster galaxies, for both sets of limits, are detectable only in the $z=3.5$--$4$ bin and only at high luminosities (\mbox{$L_{\mathrm{IR}} > 10^{12.5}\,L_{\odot}$}). The shape of the field IRLF for the EWS$+$SPIRE-detectable is unchanged with redshift, although the small area of the MAMBO lightcone limits the statistical utility of this sample. The EWS$+$PACS-detectable sample is more statistically robust, but since PACS observations to the required depth and area do not exist, we must rely on other existing FIR and (sub-)mm facilities, as well as upcoming FIR (concept) facilities such as PRIMA and CCAT.

    \item To enable a statistically robust study of the environment dependence of the IRLF for FIR- and \Euclid-detectable sources, a significantly larger simulation area than \SI{3.14}{\deg\squared} is required. Using the predictions for the EWS$+$SPIRE-detectable sample in Fig.~\ref{LF fir env comparison normalised}, we estimate that a minimum area of \SI{90}{\deg\squared} -- $30$ times the current MAMBO lightcone -- is necessary to derive statistically significant results.
\end{enumerate}

Overall, the FIR extension of the MAMBO mock successfully reproduces the expected trends of the FIR physical properties with redshift and highlights the expected flux ranges of FIR sources that we can study with real \textit{Euclid} data. The number counts and IRLFs are mostly in agreement with the various published data. Through environment-based IRLFs -- particularly with the addition of NIR/FIR detection limits -- we forecast the capabilities of future \Euclid observations in characterising galaxy environments. These results highlight the promise of combining \Euclid with current and future FIR data/facilities and emphasise the need for larger simulated and observational fields to explore the environmental evolution of IRLFs in a statistically robust manner.

\begin{acknowledgements}
  
\AckEC  
The Herschel spacecraft was designed, built, tested, and launched under a contract to ESA managed by the Herschel/Planck Project team by an industrial consortium under the overall responsibility of the prime contractor Thales Alenia Space (Cannes), and including Astrium (Friedrichshafen) responsible for the payload module and for system testing at spacecraft level, Thales Alenia Space (Turin) responsible for the service module, and Astrium (Toulouse) responsible for the telescope, with in excess of a hundred subcontractors. SPIRE has been developed by a consortium of institutes led by Cardiff University (UK) and including Univ. Lethbridge (Canada); NAOC (China); CEA, LAM (France); IFSI, Univ. Padua (Italy); IAC (Spain); Stockholm Observatory (Sweden); Imperial College London, RAL, UCL-MSSL, UKATC, Univ. Sussex (UK); and Caltech, JPL, NHSC, Univ. Colorado (USA). This development has been supported by national funding agencies: CSA (Canada); NAOC (China); CEA, CNES, CNRS (France); ASI (Italy); MCINN (Spain); SNSB (Sweden); STFC, UKSA (UK); and NASA (USA). PACS has been developed by a consortium of institutes led by MPE (Germany) and including UVIE (Austria); KU Leuven, CSL, IMEC (Belgium); CEA, LAM (France); MPIA (Germany); INAF-IFSI/OAA/OAP/OAT, LENS, SISSA (Italy); IAC (Spain). This development has been supported by the funding agencies BMVIT (Austria), ESA-PRODEX (Belgium), CEA/CNES (France), DLR (Germany), ASI/INAF (Italy), and CICYT/MCYT (Spain).

\end{acknowledgements}

\bibliography{Euclid}

\begin{appendix}

\section{Stacked FIR flux densities}

Here we show the stacked FIR flux densities of \Euclid-detectable star-forming galaxies sorted into stellar mass and redshift bins to match those in the analysis of \cite{Q1-SP071}.
Across all stellar mass bins, the FIR flux densities in each band show a systematic decline with increasing redshift, reflecting cosmological dimming. 
Comparing these average FIR flux densities with the early results of \cite{Q1-SP071}, our MAMBO predictions are generally in agreement.

\onecolumn
\begin{center}
\centering
\tablefirsthead{
\hline
\hline
 & & & & & & & \\[-8pt]
$\log_{10}$ & $z$ & $S_{100}$ & $S_{160}$ & $S_{250}$ & $S_{350}$ & $S_{500}$ & $S_{850}$ \\
$(M_*/M_\odot)$ & & [mJy] & [mJy] & [mJy] & [mJy] & [mJy] & [mJy]\\
 & & & & & & & \\[-8pt]
\hline
\hline
 & & & & & & & \\[-8pt]}
\tablehead{
\hline
\hline
 & & & & & & & \\[-8pt]
$\log_{10}$ & $z$ & $S_{100}$ & $S_{160}$ & $S_{250}$ & $S_{350}$ & $S_{500}$ & $S_{850}$ \\
$(M_*/M_\odot)$ & & [mJy] & [mJy] & [mJy] & [mJy] & [mJy] & [mJy]\\
 & & & & & & & \\[-8pt]
\hline
\hline
 & & & & & & & \\[-8pt]}
\tablecaption{Average FIR flux densities of EWS-detectable galaxies in stellar mass and redshift bins as predicted by MAMBO. We find no EWS-detectable galaxies in any of the FIR bands at $z = 2.9$ for a stellar mass of $\log_{10}(\mathit{M}_{*} / M_{\odot}) = 8.5$; this bin is therefore not shown.}
\begin{supertabular}{c|c|c|c|c|c|c|c}
\label{tab:average fluxes}
8.5 & 0.3 & 0.15$\pm$0.003 & 0.4$\pm$0.005 & 0.27$\pm$0.003 & 0.15$\pm$0.002 & 0.063$\pm$0.0007 & 0.02$\pm$0.0002 \\
8.5 & 0.5 & 0.05$\pm$0.001 & 0.2$\pm$0.002 & 0.14$\pm$0.001 & 0.08$\pm$0.0006 & 0.035$\pm$0.0003 & 0.01$\pm$0.00008 \\
8.5 & 0.7 & 0.03$\pm$0.001 & 0.1$\pm$0.001 & 0.1$\pm$0.001 & 0.062$\pm$0.0006 & 0.029$\pm$0.0003 & 0.0073$\pm$0.00006 \\
8.5 & 0.9 & 0.02$\pm$0.001 & 0.1$\pm$0.002 & 0.078$\pm$0.001 & 0.053$\pm$0.0009 & 0.025$\pm$0.0004 & 0.0066$\pm$0.0001 \\
8.5 & 1.1 & 0.02$\pm$0.001 & 0.1$\pm$0.004 & 0.068$\pm$0.003 & 0.049$\pm$0.002 & 0.023$\pm$0.0007 & 0.006$\pm$0.0002 \\
8.5 & 1.3 & 0.02$\pm$0.002 & 0.1$\pm$0.006 & 0.07$\pm$0.004 & 0.051$\pm$0.003 & 0.023$\pm$0.001 & 0.006$\pm$0.0003 \\
8.5 & 1.5 & 0.02$\pm$0.007 & 0.1$\pm$0.02 & 0.064$\pm$0.01 & 0.048$\pm$0.009 & 0.021$\pm$0.003 & 0.0055$\pm$0.0008 \\
8.5 & 1.7 & 0.01$\pm$0.005 & 0.03$\pm$0.02 & 0.029$\pm$0.01 & 0.021$\pm$0.009 & 0.014$\pm$0.006 & 0.0032$\pm$0.001 \\
8.5 & 1.9 & 0.001$\pm$0.001 & 0.01$\pm$0.002 & 0.0074$\pm$0.002 & 0.0057$\pm$0.002 & 0.0051$\pm$0.002 & 0.00096$\pm$0.0003 \\
8.5 & 2.1 & 0.001$\pm$0.001 & 0.01$\pm$0.005 & 0.0071$\pm$0.006 & 0.0056$\pm$0.004 & 0.0038$\pm$0.002 & 0.001$\pm$0.0005 \\
8.5 & 2.3 & 0.0002$\pm$0.0001 & 0.003$\pm$0.0006 & 0.005$\pm$0.001 & 0.0052$\pm$0.001 & 0.0044$\pm$0.001 & 0.0014$\pm$0.0004 \\
8.5 & 2.5 & 0.0003$\pm$0.002 & 0.002$\pm$0.002 & 0.0028$\pm$0.002 & 0.0022$\pm$0.001 & 0.0033$\pm$0.002 & 0.00036$\pm$0.0002 \\
8.5 & 2.7 & 0.0004$\pm$0.002 & 0.003$\pm$0.001 & 0.0049$\pm$0.002 & 0.0046$\pm$0.001 & 0.0041$\pm$0.0002 & 0.0011$\pm$0.0003 \\
8.9 & 0.3 & 0.44$\pm$0.01 & 1.1$\pm$0.02 & 0.77$\pm$0.01 & 0.43$\pm$0.005 & 0.18$\pm$0.002 & 0.052$\pm$0.0006 \\
8.9 & 0.5 & 0.13$\pm$0.002 & 0.4$\pm$0.004 & 0.36$\pm$0.003 & 0.22$\pm$0.002 & 0.096$\pm$0.0007 & 0.026$\pm$0.0002 \\
8.9 & 0.7 & 0.06$\pm$0.001 & 0.3$\pm$0.002 & 0.24$\pm$0.002 & 0.15$\pm$0.001 & 0.072$\pm$0.0004 & 0.018$\pm$0.0001 \\
8.9 & 0.9 & 0.04$\pm$0.001 & 0.2$\pm$0.002 & 0.18$\pm$0.001 & 0.12$\pm$0.0008 & 0.06$\pm$0.0004 & 0.016$\pm$0.0001 \\
8.9 & 1.1 & 0.03$\pm$0.001 & 0.1$\pm$0.002 & 0.14$\pm$0.001 & 0.11$\pm$0.0009 & 0.054$\pm$0.0004 & 0.015$\pm$0.0001 \\
8.9 & 1.3 & 0.02$\pm$0.001 & 0.1$\pm$0.002 & 0.12$\pm$0.002 & 0.095$\pm$0.001 & 0.049$\pm$0.0005 & 0.014$\pm$0.0001 \\
8.9 & 1.5 & 0.01$\pm$0.0004 & 0.1$\pm$0.002 & 0.097$\pm$0.002 & 0.08$\pm$0.001 & 0.043$\pm$0.0006 & 0.013$\pm$0.0002 \\
8.9 & 1.7 & 0.01$\pm$0.001 & 0.1$\pm$0.004 & 0.09$\pm$0.003 & 0.071$\pm$0.002 & 0.043$\pm$0.001 & 0.013$\pm$0.0003 \\
8.9 & 1.9 & 0.01$\pm$0.001 & 0.1$\pm$0.006 & 0.087$\pm$0.006 & 0.07$\pm$0.004 & 0.045$\pm$0.002 & 0.013$\pm$0.0006 \\
8.9 & 2.1 & 0.02 $\pm$0.007& 0.1$\pm$0.04 & 0.14$\pm$0.04 & 0.11$\pm$0.03 & 0.063$\pm$0.01 & 0.017$\pm$0.003 \\
8.9 & 2.3 & 0.004$\pm$0.001 & 0.03$\pm$0.007 & 0.043$\pm$0.009 & 0.038$\pm$0.007 & 0.027$\pm$0.005 & 0.0081$\pm$0.001 \\
8.9 & 2.5 & 0.004$\pm$0.002 & 0.03$\pm$0.01 & 0.031$\pm$0.01 & 0.026$\pm$0.008 & 0.017$\pm$0.004 & 0.0053$\pm$0.001 \\
8.9 & 2.7 & 0.001$\pm$0.001 & 0.01$\pm$0.002 & 0.019$\pm$0.003 & 0.02$\pm$0.003 & 0.017$\pm$0.002 & 0.0054$\pm$0.0007 \\
8.9 & 2.9 & 0.001$\pm$0.0001 & 0.01$\pm$0.0008 & 0.011$\pm$0.001 & 0.012$\pm$0.002 & 0.0096$\pm$0.001 & 0.0036$\pm$0.0006 \\
9.3 & 0.3 & 1.16$\pm$0.04 & 2.8$\pm$0.05 & 2$\pm$0.03 & 1.1$\pm$0.02 & 0.47$\pm$0.007 & 0.13$\pm$0.002 \\
9.3 & 0.5 & 0.39$\pm$0.01 & 1.2$\pm$0.02 & 1$\pm$0.01 & 0.6$\pm$0.006 & 0.27$\pm$0.002 & 0.071$\pm$0.0007 \\
9.3 & 0.7 & 0.17$\pm$0.003 & 0.7$\pm$0.007 & 0.64$\pm$0.005 & 0.41$\pm$0.003 & 0.19$\pm$0.001 & 0.05$\pm$0.0003 \\
9.3 & 0.9 & 0.10$\pm$0.002 & 0.5$\pm$0.004 & 0.47$\pm$0.003 & 0.33$\pm$0.002 & 0.16$\pm$0.001 & 0.044$\pm$0.0003 \\
9.3 & 1.1 & 0.06$\pm$0.001 & 0.4$\pm$0.004 & 0.38$\pm$0.003 & 0.29$\pm$0.002 & 0.15$\pm$0.001 & 0.041$\pm$0.0003 \\
9.3 & 1.3 & 0.04$\pm$0.001 & 0.3$\pm$0.003 & 0.31$\pm$0.003 & 0.25$\pm$0.002 & 0.13$\pm$0.0009 & 0.039$\pm$0.0003 \\
9.3 & 1.5 & 0.03$\pm$0.001 & 0.2$\pm$0.003 & 0.27$\pm$0.003 & 0.22$\pm$0.002 & 0.12$\pm$0.001 & 0.037$\pm$0.0003 \\
9.3 & 1.7 & 0.03$\pm$0.001 & 0.2$\pm$0.004 & 0.25$\pm$0.004 & 0.2$\pm$0.002 & 0.13$\pm$0.001 & 0.038$\pm$0.0004 \\
9.3 & 1.9 & 0.02$\pm$0.001 & 0.2$\pm$0.005 & 0.25$\pm$0.006 & 0.21$\pm$0.004 & 0.13$\pm$0.002 & 0.041$\pm$0.0007 \\
9.3 & 2.1 & 0.02$\pm$0.001 & 0.2$\pm$0.008 & 0.23$\pm$0.009 & 0.2$\pm$0.007 & 0.13$\pm$0.004 & 0.04$\pm$0.001 \\
9.3 & 2.3 & 0.02$\pm$0.002 & 0.1$\pm$0.01 & 0.19$\pm$0.01 & 0.17$\pm$0.009 & 0.11$\pm$0.005 & 0.037$\pm$0.002 \\
9.3 & 2.5 & 0.01$\pm$0.001 & 0.1$\pm$0.007 & 0.13$\pm$0.009 & 0.13$\pm$0.007 & 0.088$\pm$0.005 & 0.03$\pm$0.001 \\
9.3 & 2.7 & 0.01$\pm$0.0004 & 0.1$\pm$0.004 & 0.1$\pm$0.007 & 0.11$\pm$0.007 & 0.081$\pm$0.005 & 0.03$\pm$0.002 \\
9.3 & 2.9 & 0.01$\pm$0.001 & 0.1$\pm$0.006 & 0.11$\pm$0.01 & 0.11$\pm$0.009 & 0.081$\pm$0.006 & 0.031$\pm$0.002 \\
9.7 & 0.3 & 2.69$\pm$0.1 & 6.5$\pm$0.2 & 4.8$\pm$0.1 & 2.7$\pm$0.05 & 1.1$\pm$0.02 & 0.31$\pm$0.006 \\
9.7 & 0.5 & 1.02$\pm$0.04 & 3.1$\pm$0.05 & 2.6$\pm$0.04 & 1.5$\pm$0.02 & 0.67$\pm$0.008 & 0.17$\pm$0.002 \\
9.7 & 0.7 & 0.45$\pm$0.01 & 1.8$\pm$0.02 & 1.7$\pm$0.02 & 1.1$\pm$0.009 & 0.51$\pm$0.004 & 0.13$\pm$0.001 \\
9.7 & 0.9 & 0.26$\pm$0.005 & 1.3$\pm$0.01 & 1.3$\pm$0.01 & 0.89$\pm$0.007 & 0.44$\pm$0.003 & 0.12$\pm$0.0008 \\
9.7 & 1.1 & 0.18$\pm$0.004 & 1$\pm$0.01 & 1$\pm$0.01 & 0.77$\pm$0.006 & 0.4$\pm$0.003 & 0.11$\pm$0.0008 \\
9.7 & 1.3 & 0.12$\pm$0.003 & 0.8$\pm$0.01 & 0.88$\pm$0.009 & 0.7$\pm$0.006 & 0.37$\pm$0.003 & 0.11$\pm$0.0008 \\
9.7 & 1.5 & 0.09$\pm$0.002 & 0.6$\pm$0.008 & 0.76$\pm$0.008 & 0.62$\pm$0.005 & 0.35$\pm$0.003 & 0.11$\pm$0.0008 \\
9.7 & 1.7 & 0.07$\pm$0.002 & 0.5$\pm$0.009 & 0.72$\pm$0.009 & 0.59$\pm$0.006 & 0.36$\pm$0.003 & 0.11$\pm$0.001 \\
\multicolumn{8}{r}{Continued on next page}\\
9.7 & 1.9 & 0.06$\pm$0.002 & 0.5$\pm$0.01 & 0.66$\pm$0.01 & 0.56$\pm$0.008 & 0.36$\pm$0.004 & 0.11$\pm$0.001 \\
9.7 & 2.1 & 0.06$\pm$0.002 & 0.5$\pm$0.01 & 0.66$\pm$0.01 & 0.57$\pm$0.01 & 0.37$\pm$0.006 & 0.12$\pm$0.002 \\
9.7 & 2.3 & 0.05$\pm$0.002 & 0.4$\pm$0.02 & 0.61$\pm$0.02 & 0.56$\pm$0.01 & 0.37$\pm$0.009 & 0.13$\pm$0.003 \\
9.7 & 2.5 & 0.03$\pm$0.002 & 0.3$\pm$0.02 & 0.49$\pm$0.02 & 0.48$\pm$0.02 & 0.33$\pm$0.01 & 0.12$\pm$0.004 \\
9.7 & 2.7 & 0.03$\pm$0.002 & 0.3$\pm$0.01 & 0.44$\pm$0.02 & 0.44$\pm$0.02 & 0.31$\pm$0.01 & 0.12$\pm$0.004 \\
9.7 & 2.9 & 0.02$\pm$0.001 & 0.2$\pm$0.01 & 0.36$\pm$0.02 & 0.38$\pm$0.02 & 0.27$\pm$0.01 & 0.1$\pm$0.004 \\
10.1 & 0.3 & 6.58$\pm$0.5 & 15$\pm$0.5 & 11$\pm$0.3 & 5.9$\pm$0.1 & 2.5$\pm$0.05 & 0.64$\pm$0.01 \\
10.1 & 0.5 & 2.40$\pm$0.09 & 7.4$\pm$0.1 & 6.1$\pm$0.1 & 3.6$\pm$0.05 & 1.6$\pm$0.02 & 0.41$\pm$0.006 \\
10.1 & 0.7 & 1.16$\pm$0.05 & 4.6$\pm$0.08 & 4.2$\pm$0.06 & 2.7$\pm$0.03 & 1.3$\pm$0.01 & 0.33$\pm$0.003 \\
10.1 & 0.9 & 0.75$\pm$0.03 & 3.5$\pm$0.05 & 3.4$\pm$0.04 & 2.4$\pm$0.02 & 1.2$\pm$0.01 & 0.31$\pm$0.003 \\
10.1 & 1.1 & 0.46$\pm$0.02 & 2.5$\pm$0.04 & 2.7$\pm$0.03 & 2.1$\pm$0.02 & 1.1$\pm$0.01 & 0.3$\pm$0.003 \\
10.1 & 1.3 & 0.34$\pm$0.02 & 2$\pm$0.03 & 2.4$\pm$0.03 & 1.9$\pm$0.02 & 1$\pm$0.009 & 0.3$\pm$0.003 \\
10.1 & 1.5 & 0.27$\pm$0.02 & 1.8$\pm$0.03 & 2.2$\pm$0.03 & 1.8$\pm$0.02 & 1$\pm$0.01 & 0.31$\pm$0.003 \\
10.1 & 1.7 & 0.21$\pm$0.01 & 1.5$\pm$0.03 & 2$\pm$0.03 & 1.6$\pm$0.02 & 1$\pm$0.01 & 0.31$\pm$0.003 \\
10.1 & 1.9 & 0.16$\pm$0.01 & 1.3$\pm$0.03 & 1.8$\pm$0.03 & 1.5$\pm$0.02 & 0.97$\pm$0.01 & 0.31$\pm$0.004 \\
10.1 & 2.1 & 0.15$\pm$0.01 & 1.2$\pm$0.03 & 1.7$\pm$0.04 & 1.5$\pm$0.03 & 1$\pm$0.02 & 0.33$\pm$0.005 \\
10.1 & 2.3 & 0.11$\pm$0.01 & 1$\pm$0.03 & 1.5$\pm$0.04 & 1.4$\pm$0.03 & 0.93$\pm$0.02 & 0.32$\pm$0.006 \\
10.1 & 2.5 & 0.09$\pm$0.01 & 0.8$\pm$0.03 & 1.4$\pm$0.04 & 1.3$\pm$0.04 & 0.92$\pm$0.02 & 0.33$\pm$0.008 \\
10.1 & 2.7 & 0.09$\pm$0.01 & 0.8$\pm$0.03 & 1.3$\pm$0.04 & 1.3$\pm$0.04 & 0.93$\pm$0.02 & 0.34$\pm$0.008 \\
10.1 & 2.9 & 0.07$\pm$0.02 & 0.7$\pm$0.04 & 1.2$\pm$0.06 & 1.3$\pm$0.05 & 0.9$\pm$0.03 & 0.35$\pm$0.01 \\
10.5 & 0.3 & 9.22$\pm$0.7 & 22$\pm$0.9 & 17$\pm$0.5 & 9.3$\pm$0.3 & 3.9$\pm$0.1 & 1$\pm$0.03 \\
10.5 & 0.5 & 3.86$\pm$0.2 & 13$\pm$0.3 & 11$\pm$0.2 & 6.4$\pm$0.1 & 2.9$\pm$0.05 & 0.72$\pm$0.01 \\
10.5 & 0.7 & 2.39$\pm$0.1 & 9.3$\pm$0.2 & 8.5$\pm$0.1 & 5.5$\pm$0.08 & 2.6$\pm$0.04 & 0.65$\pm$0.009 \\
10.5 & 0.9 & 1.54$\pm$0.06 & 7.1$\pm$0.1 & 6.9$\pm$0.1 & 4.8$\pm$0.06 & 2.4$\pm$0.03 & 0.63$\pm$0.008 \\
10.5 & 1.1 & 1.02$\pm$0.03 & 5.7$\pm$0.1 & 6$\pm$0.1 & 4.6$\pm$0.07 & 2.4$\pm$0.03 & 0.66$\pm$0.009 \\
10.5 & 1.3 & 0.77$\pm$0.02 & 4.9$\pm$0.09 & 5.8$\pm$0.09 & 4.6$\pm$0.06 & 2.5$\pm$0.03 & 0.73$\pm$0.008 \\
10.5 & 1.5 & 0.62$\pm$0.02 & 4.2$\pm$0.09 & 5.2$\pm$0.08 & 4.3$\pm$0.06 & 2.4$\pm$0.03 & 0.74$\pm$0.009 \\
10.5 & 1.7 & 0.52$\pm$0.02 & 3.7$\pm$0.09 & 4.9$\pm$0.08 & 4.1$\pm$0.06 & 2.5$\pm$0.03 & 0.78$\pm$0.009 \\
10.5 & 1.9 & 0.41$\pm$0.02 & 3.3$\pm$0.08 & 4.7$\pm$0.09 & 4.1$\pm$0.07 & 2.6$\pm$0.04 & 0.84$\pm$0.01 \\
10.5 & 2.1 & 0.38$\pm$0.02 & 3.2$\pm$0.1 & 4.6$\pm$0.1 & 4.1$\pm$0.08 & 2.7$\pm$0.05 & 0.88$\pm$0.01 \\
10.5 & 2.3 & 0.32$\pm$0.02 & 2.8$\pm$0.1 & 4.2$\pm$0.1 & 3.9$\pm$0.1 & 2.6$\pm$0.06 & 0.89$\pm$0.02 \\
10.5 & 2.5 & 0.25$\pm$0.01 & 2.4$\pm$0.1 & 3.8$\pm$0.2 & 3.7$\pm$0.2 & 2.5$\pm$0.1 & 0.9$\pm$0.03 \\
10.5 & 2.7 & 0.21$\pm$0.01 & 2$\pm$0.1 & 3.3$\pm$0.1 & 3.3$\pm$0.1 & 2.3$\pm$0.09 & 0.86$\pm$0.03 \\
10.5 & 2.9 & 0.20$\pm$0.01 & 1.9$\pm$0.1 & 3.3$\pm$0.2 & 3.4$\pm$0.2 & 2.4$\pm$0.1 & 0.91$\pm$0.04 \\
10.9 & 0.3 & 10.37$\pm$0.9 & 28$\pm$1 & 21$\pm$0.8 & 12$\pm$0.4 & 5.2$\pm$0.2 & 1.3$\pm$0.05 \\
10.9 & 0.5 & 5.27$\pm$0.4 & 18$\pm$0.6 & 15$\pm$0.4 & 9.3$\pm$0.2 & 4.2$\pm$0.09 & 1.1$\pm$0.02 \\
10.9 & 0.7 & 3.34$\pm$0.3 & 13$\pm$0.5 & 13$\pm$0.4 & 8.3$\pm$0.2 & 4$\pm$0.09 & 1$\pm$0.02 \\
10.9 & 0.9 & 2.17$\pm$0.1 & 11$\pm$0.4 & 11$\pm$0.3 & 8.2$\pm$0.2 & 4.2$\pm$0.08 & 1.1$\pm$0.02 \\
10.9 & 1.1 & 1.73$\pm$0.05 & 9.9$\pm$0.3 & 11$\pm$0.3 & 8.1$\pm$0.2 & 4.3$\pm$0.09 & 1.2$\pm$0.02 \\
10.9 & 1.3 & 1.30$\pm$0.05 & 8.5$\pm$0.2 & 10$\pm$0.2 & 8.1$\pm$0.2 & 4.4$\pm$0.08 & 1.3$\pm$0.02 \\
10.9 & 1.5 & 1.19$\pm$0.04 & 8.3$\pm$0.2 & 10$\pm$0.2 & 8.6$\pm$0.2 & 4.9$\pm$0.09 & 1.5$\pm$0.03 \\
10.9 & 1.7 & 0.98$\pm$0.03 & 7.2$\pm$0.2 & 9.7$\pm$0.2 & 8.1$\pm$0.1 & 5$\pm$0.08 & 1.6$\pm$0.03 \\
10.9 & 1.9 & 0.77$\pm$0.04 & 6.3$\pm$0.2 & 9.1$\pm$0.2 & 7.9$\pm$0.2 & 5.1$\pm$0.1 & 1.6$\pm$0.03 \\
10.9 & 2.1 & 0.75$\pm$0.04 & 6.2$\pm$0.2 & 9.3$\pm$0.3 & 8.4$\pm$0.2 & 5.5$\pm$0.1 & 1.9$\pm$0.04 \\
10.9 & 2.3 & 0.67$\pm$0.03 & 6$\pm$0.3 & 9.2$\pm$0.3 & 8.5$\pm$0.3 & 5.7$\pm$0.2 & 2$\pm$0.06 \\
10.9 & 2.5 & 0.49$\pm$0.03 & 4.5$\pm$0.3 & 7.5$\pm$0.4 & 7.5$\pm$0.3 & 5.2$\pm$0.2 & 1.9$\pm$0.08 \\
10.9 & 2.7 & 0.41$\pm$0.03 & 3.9$\pm$0.3 & 6.8$\pm$0.4 & 7.1$\pm$0.4 & 5.1$\pm$0.2 & 1.9$\pm$0.08 \\
10.9 & 2.9 & 0.35$\pm$0.03 & 3.4$\pm$0.3 & 5.9$\pm$0.5 & 6.4$\pm$0.4 & 4.6$\pm$0.3 & 1.8$\pm$0.1 \\
11.3 & 0.3 & 4.71$\pm$0.8 & 19$\pm$2 & 17$\pm$2 & 10$\pm$1 & 4.6$\pm$0.5 & 1.3$\pm$0.1 \\
11.3 & 0.5 & 4.43$\pm$0.6 & 18$\pm$2 & 17$\pm$1 & 10$\pm$0.6 & 4.8$\pm$0.3 & 1.3$\pm$0.07 \\
11.3 & 0.7 & 4.03$\pm$0.7 & 15$\pm$1 & 14$\pm$1 & 9.3$\pm$0.6 & 4.5$\pm$0.3 & 1.2$\pm$0.07 \\
11.3 & 0.9 & 2.05$\pm$0.2 & 11$\pm$0.7 & 12$\pm$0.6 & 9.2$\pm$0.4 & 4.8$\pm$0.2 & 1.4$\pm$0.06 \\
11.3 & 1.1 & 1.71$\pm$0.2 & 11$\pm$0.7 & 13$\pm$0.6 & 10$\pm$0.4 & 5.4$\pm$0.2 & 1.6$\pm$0.06 \\
\multicolumn{8}{r}{Continued on next page}\\
11.3 & 1.3 & 1.89$\pm$0.2 & 12$\pm$0.8 & 14$\pm$0.7 & 11$\pm$0.4 & 6$\pm$0.2 & 1.8$\pm$0.06 \\
11.3 & 1.5 & 1.56$\pm$0.1 & 11$\pm$0.7 & 14$\pm$0.6 & 11$\pm$0.4 & 6.6$\pm$0.2 & 2$\pm$0.06 \\
11.3 & 1.7 & 1.29$\pm$0.1 & 9.9$\pm$0.5 & 14$\pm$0.5 & 12$\pm$0.4 & 7.2$\pm$0.2 & 2.3$\pm$0.07 \\
11.3 & 1.9 & 1.39$\pm$0.1 & 11$\pm$0.8 & 15$\pm$0.8 & 13$\pm$0.6 & 8.1$\pm$0.3 & 2.6$\pm$0.09 \\
11.3 & 2.1 & 1.00$\pm$0.1 & 8.6$\pm$0.5 & 13$\pm$0.6 & 12$\pm$0.5 & 8$\pm$0.3 & 2.8$\pm$0.1 \\
11.3 & 2.3 & 1.39$\pm$0.2 & 12$\pm$1 & 17$\pm$2 & 15$\pm$1 & 10$\pm$0.6 & 3.3$\pm$0.2 \\
11.3 & 2.5 & 1.04$\pm$0.1 & 9.8$\pm$1 & 16$\pm$2 & 15$\pm$2 & 11$\pm$1 & 3.8$\pm$0.3 \\
11.3 & 2.7 & 1.07$\pm$0.3 & 9.9$\pm$2 & 16$\pm$3 & 15$\pm$2 & 10$\pm$1 & 3.7$\pm$0.4 \\
11.3 & 2.9 & 0.54$\pm$0.1 & 5.1$\pm$0.7 & 9.5$\pm$1 & 11$\pm$1 & 8.1$\pm$0.7 & 3.3$\pm$0.3\\
\hline\hline
\end{supertabular}
\end{center}
\twocolumn

\section{Integral number counts} \label{int counts}

\begin{figure*}
    \centering
    \includegraphics[width=\linewidth]{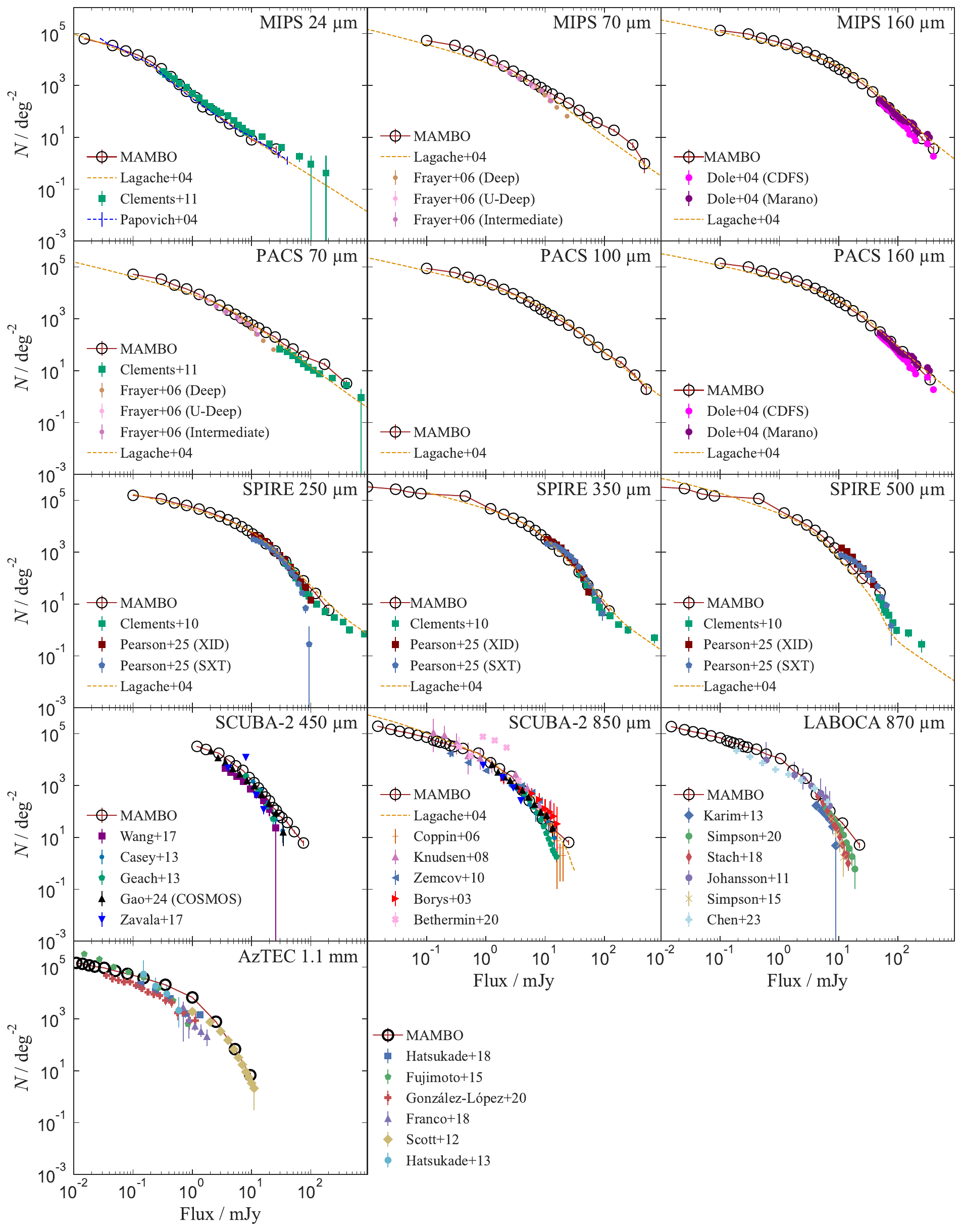}
    \caption{Integral counts of the MAMBO mock for various bands: MIPS, PACS, SPIRE, SCUBA-2, Apex LABOCA, and AzTEC. Data and models are shown for comparison.}
    \label{fig:int counts all}
\end{figure*}

\begin{figure*}
     \centering
\includegraphics[width=2\columnwidth]{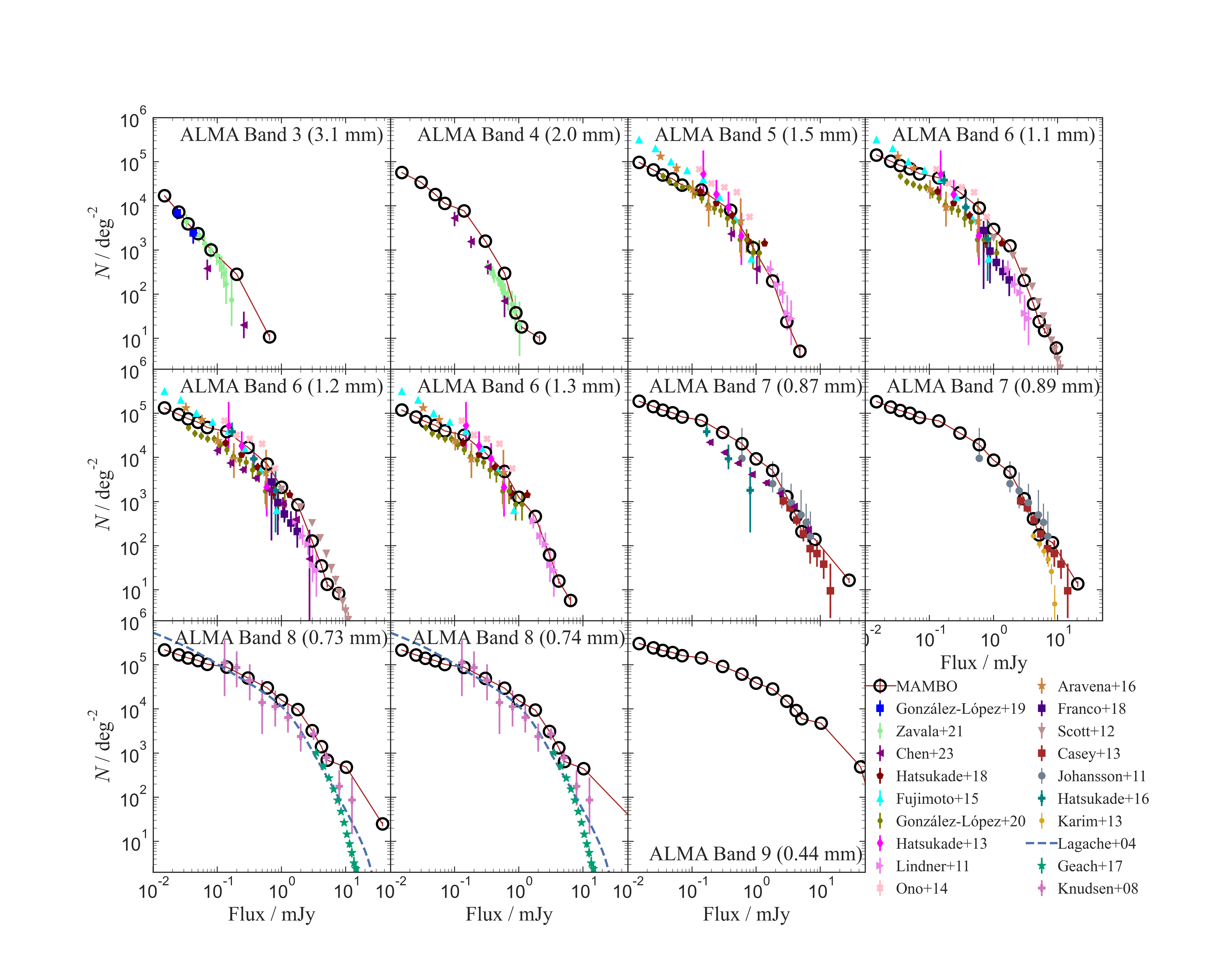}
    \caption{Integral number counts of the MAMBO mock for various ALMA bands. The specific wavelength of each ALMA band is stated in each subplot. Literature counts and models are shown for comparison.}

    \label{alma int counts}
\end{figure*}

Here, we show the integral number counts for all the bands previously mentioned in Sect.~\ref{sc:Number counts}. The integral counts predicted from MAMBO are in overall agreement with the published estimates, although there are a few cases of discrepancies. Similarly to the differential counts, for MIPS and PACS at $70\,\micron$ (Fig.~\ref{fig:int counts all}), MAMBO overestimates the integral counts at about $100\,$mJy. The MAMBO counts for rest of the MIPS and PACS bands agree well with published data. The \Herschel, SCUBA-2, and LABOCA predictions from MAMBO are generally in agreement with published results. An overprediction is seen for the AzTEC $1.1\,$mm counts where MAMBO predicts an excess at about $1\,$mJy, with a maximum discrepancy of $5\sigma$ from the published estimates. In the integral counts for ALMA Band~$8$ (Fig.~\ref{alma int counts}), the MAMBO counts deviate by more than $10\sigma$ from the literature values (when compared to \citealt{Geach_2017}) at fluxes $\gtrsim 5\,$mJy but we note that this is likely due to comparing Band 8 data with higher wavelength ($\sim 850\,\micron$) published data.

\end{appendix}

\end{document}